\documentclass[aps,prb,nofootinbib,twocolumn,balancelastpage,amsmath,amssymb,floatfix,superscriptaddress,]{revtex4-1}

\usepackage{xcolor}

\usepackage{bm}
\usepackage{mathtools}
\usepackage{subfigure}

\newcommand{\tr}{\operatorname{Tr}}
\newcommand{\ketbra}[2]{\ket {#1} \hskip -0.8ex \bra {#2}}
\newcommand{\expct}[1]{\left\langle #1 \right\rangle}

\newcommand{\graph}{\mathrm{graph}}

\usepackage{amssymb}
\usepackage{amsmath}
\usepackage{amsfonts}
\usepackage{braket}
\usepackage{graphicx}
\usepackage{epstopdf}
\usepackage[colorlinks,linkcolor=blue]{hyperref}
\usepackage{epsfig}
\usepackage[toc,page,title,titletoc,header]{appendix}
\usepackage{dsfont,amsthm,amsbsy}
\usepackage{hyperref}
\usepackage[xspace]{ellipsis}
\usepackage{cancel}

\DeclareMathOperator{\supp}{supp}
\DeclareMathOperator{\diam}{diam}
\DeclareMathOperator*{\mean}{mean}

\newcommand{\pool}{\mathcal G}
\newcommand{\semiemp}{\text{semi-emp.}}
\newcommand{\Th}{{\mathrm{Th}}}
\newcommand{\Haar}{\text{Haar}}

\begin{document}
\title{Effective dissipation rate in a Liouvillian graph picture of high-temperature quantum hydrodynamics}% in strongly-interacting spin systems}

\author{Christopher David White}
\affiliation{Joint Center for Quantum Information and Computer Science, NIST/University of Maryland, College Park, Maryland 20742, USA}
\affiliation{Condensed Matter Theory Center, University of Maryland, College Park, Md, 20742}
\email{cdwhite@umd.edu}

\begin{abstract}

  At high temperature, generic strongly interacting spin systems are expected to display hydrodynamics:
  local transport of conserved quantities,
  governed by classical partial differential equations like the diffusion equation.
  %This process is well-understood in $U(1)$ conserving random spin systems,
  %Practical methods justified by this intuition.
  %%%%%%%
  I argue that the emergence of this dissipative long-wavelength dynamics from the system's unitary microscopic dynamics 
  is controlled by the structure of the \textit{Liouvillian graph} of the system's Hamiltonian,
  that is, the graph induced on Pauli strings by commutation with that Hamiltonian.
  The Liouvillian graph decomposes naturally into subgraphs of Pauli strings of constant diameter,
  and the coherent dynamics of these subgraphs determines the rate at which operator weight spreads to long operators.
  %%%%%%%
  This argument provides a quantitative theory of the emergence of a dissipative effective dynamics from unitary microscopic dynamics;
  it also leads to an effective model with Hilbert space dimension linear in system size and exponential in the UV cutoff for diffusion.

\end{abstract}
  
%Generic interacting quantum systems are expected to relax to local equilibrium by local transport of conserved quantities like energy and particle number (``hydrodynamics'').
%Two recent numerical methods motivated by this expectation have found some success \cite{}\todo{} by exactly preserving local quantities, while truncating long-range correlations.

%I justify these methods in the case of the longitudinal-field Ising model
%as perturbations around a so-called ``operator graph'' picture of hydrodynamics.
%I first write the Heisenberg dynamics of Pauli strings as single-particle hopping on a graph (the model's ``operator graph'').
%The operator graph decomposes into subgraphs consisting of Pauli strings of fixed length;
% these subgraphs are tightly intraconnected but loosely interconnected.
% I argue via a combination of numerics and analytics
% that this subgraph decomposition leads to an effective dissipation on long length scales
% that mirrors the truncation performed by the successful algorithms.
% I then give an exactly solvable effective model for diffusion in strongly-interacting quantum systems,
% and discuss UV cutoffs in space and time for diffusive hydrodynamics.
% Time permitting, I will speculate about implications for the effectiveness of numerical methods for quantum hydrodynamics,
% and about potential short-length-scale phenomenologies for systems displaying diffusion on long length scales.

%\end{abstract}
\maketitle

\section{Introduction}

Generic interacting quantum systems are expected to relax to local equilibrium by local transport of conserved quantities like energy and particle number.
This effective dynamics is called ``hydrodynamics'';
for lattice systems, in which continuous translation symmetry is broken,
it predicts at leading order diffusive behavior
\begin{align}
  \label{eq:diff}
  \frac{\partial}{\partial t} \varepsilon(x,t) = D \nabla^2 \varepsilon(x,t)\;,
\end{align}
where $\varepsilon$ is a conserved quantity like energy density
\cite{forster_hydrodynamic_1975,chaikin_principles_2000}%
.\footnote{
  Higher-order corrections to the diffusion equation \eqref{eq:diff}
  can give nontrivial physics,
  including ``long-time tails''\cite{kovtun_hydrodynamic_2003,mukerjee_towards_2006,spohn_nonlinear_2014,lux_hydrodynamic_2014,kirkpatrick_non-hydrodynamic_2021}.
  This work treats the microscopic physics leading to the diffusion equation (with or without corrections) rather than the physics described by that equation---but the interplay between that microscopic physics
  and the long-time tails,
  analogous to the interplay investigated for the Boltzmann equation in \onlinecite{kirkpatrick_non-hydrodynamic_2021}, may be an interesting future line of inquiry.
}
The emergence of hydrodynamics from microscopic unitary quantum dynamics is not well understood.
What approximations lead to efficient, accurate classical simulations of systems displaying hydrodynamics?
What are the UV cutoffs for diffusion in space and time---that is, the length- and time-scales below which the diffusion equation \eqref{eq:diff} fails to accurately describe the system's dynamics?
And when and how does hydrodynamics break down?
A theory of how hydrodynamics emerges from generic unitary dynamics will bear on many other open questions, from the nature of the many body localization transition
\footnote{Our understanding of the many body localization transition rests on phenomenological renormalization group theories that take for granted the emergence of hydrodynamics from sufficiently strongly-interacting systems.}
\cite{potter_universal_2015,vosk_theory_2015,dumitrescu_scaling_2017,goremykina_analytically_2018,morningstar_renormalization-group_2019,morningstar_many-body_2020,de_roeck_stability_2017,thiery_many-body_2017,thiery_microscopically_2017}
to experimental studies of the phase diagram of QCD
\footnote{Heavy-ion collisions produce high-temperature clouds of quarks and gluons.
  As these clouds expand and cool, they are pass through an expected line of first-order phase transitions that terminate in a ``critical end point''.
  Statistical fluctuation ``freeze out'' via the Kibble-Zurek mechanism when the cloud passes near the critical endpoint,
  leading to experimentally visible signatures of the details of the transition.
  But detailed comparison of experiments to theory requires first-principles calculations of hydrodynamical transport coefficients.
}
\cite{stephanov_signatures_1998,kolb_anisotropic_2000,ollitrault_anisotropy_1992,star_collaboration_experimental_2010,heinz_exploring_2015}
.

Many lines of evidence suggest that for generic spin systems%
---that is, systems not close to any kind of integrability or localization---%
dissipative hydrodynamics results from an insensitivity of local dynamics to long-range correlations.
As the system evolves it develops correlations.
These correlations can of course be ignored when computing local expectation values given the state at some time,
but they can also---the hypothesis runs---be ignored in estimating the system's dynamics.
This is similar in spirit to the eigenstate thermalization hypothesis,
which explains how systems can thermalize on timescales much shorter than the level spacing time
by positing that eigenstates have local expectation values very close to those of a Gibbs state \cite{deutsch_quantum_1991,srednicki_chaos_1994,srednicki_chaos_1994,dalessio_quantum_2016}.

The hypothesis that in thermalizing systems
the dynamics of local expectation values is insensitive to that of long-range correlations
intuitively justified
two matrix product operator methods, ``density matrix truncation'' (DMT) \cite{white_quantum_2018,ye_emergent_2019} and ``dissipation-assisted operator evolution'' (DAOE) \cite{rakovszky_dissipation-assisted_2020}.
Each method reduces the magnitude of the long-ranged correlations in its own way,
giving the requisite nonunitarity.
The success of these methods
provides tentative evidence that
correlations longer than a few lattice spacings can indeed be ignored on timescales longer than some cutoff time of order the coupling time.
This insensitivity hypothesis is strengthened by exact calculations of operator spreading in random unitary circuits \cite{ho_entanglement_2017,von_keyserlingk_operator_2018,nahum_quantum_2017,khemani_operator_2018,rakovszky_diffusive_2017}; 
\onlinecite{khemani_operator_2018} elegantly explained the effective nonunitarity of diffusive dynamics in $U(1)$-conserving random matrices
as the emission of non-conserved operators,
which spread ballistically,
from a diffusively relaxing ``lump'' of conserved charge.
More recent work interprets the hypothesis as a flow of information to long operators.\cite{kvorning_time-evolution_2021}

Meanwhile \onlinecite{altman_ehud_computing_2018} (cf \onlinecite{parker_universal_2019,cao_statistical_2021}),
motivated in part by work on random unitaries,
gave an effective model for the Heisenberg-picture generated by a spin Hamiltonian.
To write this effective model they mapped the Heisenberg dynamics
of some initial Pauli string $\sigma^{\bm \mu}$
to single-particle dynamics on
the graph resulting from successive commutations with the Hamiltonian;
this graph is called the \textit{Liouvillian graph}.
They then truncated this dynamics
by dropping operators farther than some distance (e.g. Hamming distance)
from the initial operator
and endowing Pauli strings at the boundary with an artificial decay.

The resulting effective model succinctly articulates the physical intuition behind DAOE and DMT,
and it translates the result of \onlinecite{khemani_operator_2018} from random unitary dynamics to Hamiltonian dynamics
(though it assumes, rather than justifying, that result).
Most enticingly,
it works directly in terms of the operator Hilbert space,
rather than matrix product operators.
So---unlike DMT and DAOE---to the extent it reflects the physical processes giving rise to diffusion,
it promises to work in more than one dimension.

But the effective model leaves open many important questions.
%At what Hamming distance or length scale can one truncate%
%---that is, at what length scale does the effective model match the dynamics of the original Hamiltonian?
In particular, its dynamics depend crucially on the magnitude of the decay imposed on boundary operators,
which is a free parameter.
%how can one justify the existence of the decay, or predict its magnitude?
This decay rate is of more than passing interest:
it is not only required for effective model to be useful for numerical experiments,
but also important for judging the accuracy of DMT and DAOE simulations, because%
---to the extent that DMT and DAOE artificially impose such a decay---%
one wishes that artificial decay to be small compared to the physical decay rate captured by the effective model.
Moreover, the decay rate is a clue to UV cutoff time for diffusion:
that is, the timescale below which coherent quantum dynamics important.

I argue that the structure of the Liouvillian graph controls the decay rate.
In particular,
if one organizes the Liouvillian graph not by distance from an initial Pauli string
but rather by Pauli string diameter
(that is, the size of the region on which a given Pauli string acts nontrivially)
one finds that operators of some diameter $l$
have many more connections to other operators of diameter $l$
than to operators of diameter $l+1$,
and more connections to operators of diameter $l+1$
than to operators of diameter $l-1$.
Operator weight that reaches the thickly connected pool of diameter $l$, then,
quickly spreads through that pool
and then to longer operators.
Via a quantum Zeno like effect,
the rate of this spread controls
the rate at which operator weight escapes into the pool of length-$l$ operators in the first place.
We can understand this spreading as a decay from---an imaginary self-energy for---the initial length-$l$ operator.
The effective model, then,
consists of treating operators with diameter $l \le l_*$ exactly,
and replacing the dynamics of operators with diameter $l > l_*$ by a decay
with decay rate given by the rate of spreading in pools of diameter $l_*+1$ and $l_* + 2$.
With this decay rate the effective model predicts diffusion coefficients close to those found by numerical simulation of the full dynamics.

In Sec.~\ref{s:model} I describe the models I treat and the quantities I measure.
In Sec.~\ref{s:op-graph} I define the Liouvillian graph and describe its structure.
In Sec.~\ref{s:rate-loss} I use that structure to predict the decay rate, and hence the rate at which operator weight escapes from short operators.
In Sec.~\ref{s:eff-model} I explicitly give my effective model,
which is similar but not identical to that of \onlinecite{altman_ehud_computing_2018},
and compare the diffusion coefficient in the effective model with the predicted decay rate
to that of the full dynamics.
I conclude in Sec.~\ref{s:discussion} by discussing implications and limitations of this Liouvillian graph picture
and the resulting effective model.

\section{Models and quantities of interest}\label{s:model}

\begin{figure*}[t]
  \begin{minipage}{0.4\textwidth}
    \includegraphics[width=\textwidth]{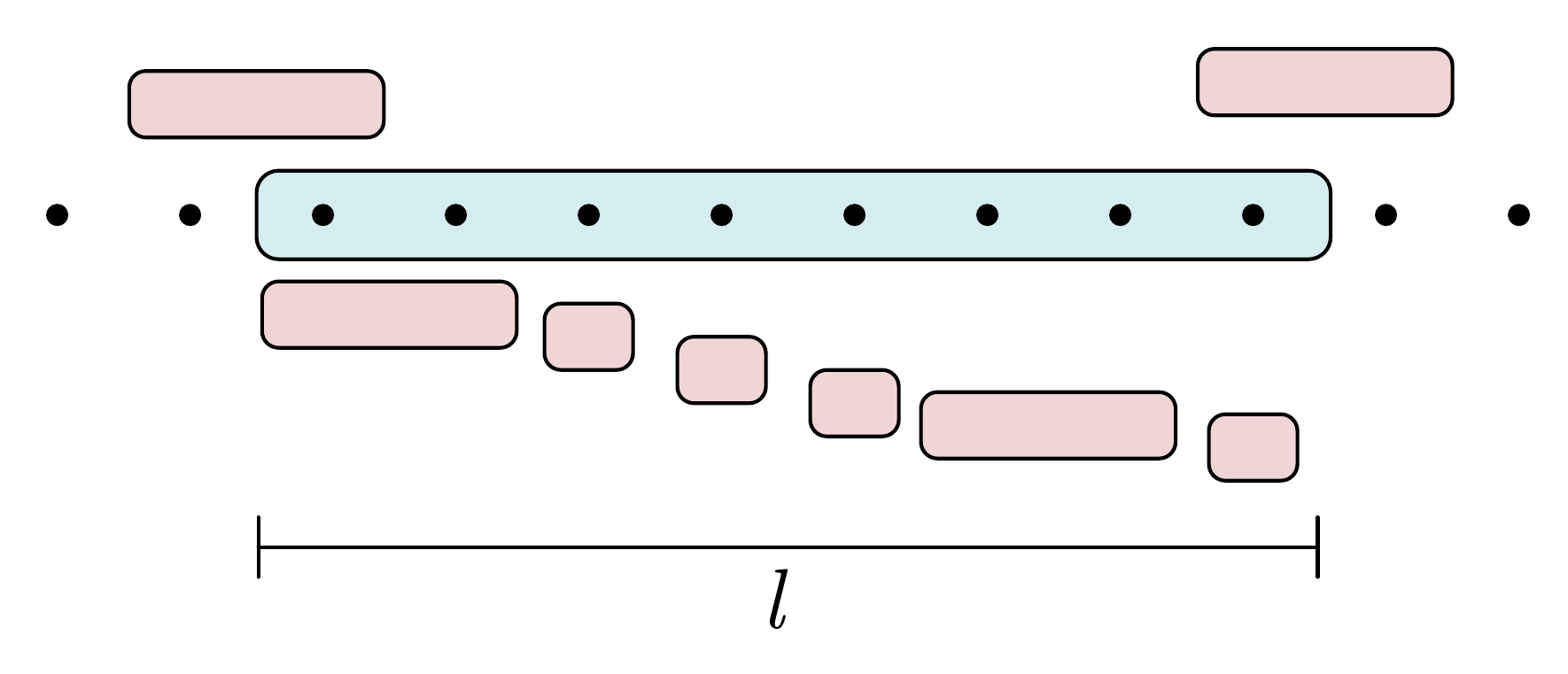}
  \end{minipage}\hskip 1cm
  \begin{minipage}{0.5\textwidth}
    \includegraphics[width=\textwidth]{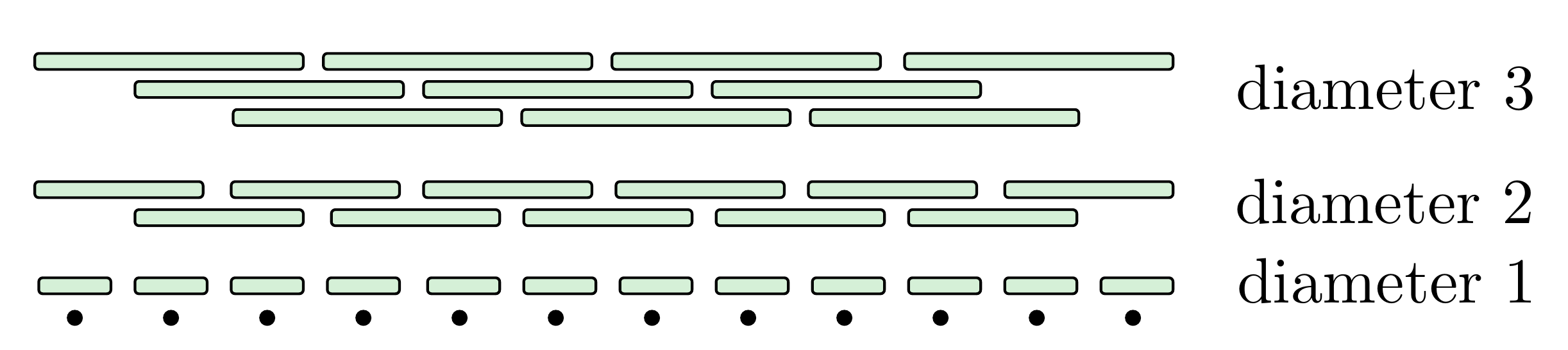}
  \end{minipage}
  \caption{\textbf{Structure of the Liouvillian graph}.
    \textbf{Left:}
    Hamiltonian terms (red) that act upon
    a Pauli string $\sigma^{\bm \mu}$ (blue) of diameter $l$.
    Commutation with any of the $O(l)$ Hamiltonian terms that completely overlap $\sigma^{\bm \mu}$ (illustrated below $\sigma^{\bm \mu}$) results in another string of diameter $l$;
    commutation with one of the $O(1)$ two-site terms overlapping on a single site
    (illustrated above $\sigma^{\bm \mu}$ results in a longer operator.
    \textbf{Right:} the resulting structure of thickly-intraconnected subgraphs (green).
    Each subgraph (``pool'') $\pool_l$ of diameter $l$ is connected to two subgraphs $\pool_{l-1}$  and two subgraphs $\pool_{l+1}$, but not (directly) to any other subgraph $\pool_l$.
  }
  \label{fig:structure}
\end{figure*}

I consider nearest-neighbor Hamiltonians on spin-1/2 chains: that is, Hamiltonians of the form
\begin{align}
  \label{eq:nn-ham}
  H = \sum_{j} \varepsilon_j,\quad \varepsilon_j = \sum_{\alpha\beta} J^{(j)}_{\alpha\beta}\sigma_j^\alpha \sigma_{j+1}^\beta\;.
\end{align}
where $\sigma^\alpha$ are the Pauli matrices, $\sigma^0$ the identity.
I take all the nonzero $J^{\alpha\beta}_k$ to have comparable magnitude,
and I restrict myself to systems that display neither integrability nor many-body localization.
My arguments apply, \textit{mutatis mutandis}, to chains of any finite onsite Hilbert space dimension, with any choice of basis for the space of onsite operators.

For a concrete example I use the transverse-field Ising model with longitudinal field on a 1D chain with periodic boundary conditions:
\begin{equation}
  \label{eq:tfim}
  H =  J\sum_{j = 1}^{L} \sigma ^z_j \sigma ^z_{j+1} + h_x\sum_j \sigma ^x_j + h_z\sum_j \sigma ^z_j\;,
\end{equation}
taking $L+1 \equiv 1$.
I write an energy density on bonds
\begin{align}
  \varepsilon_j = \sigma ^z_j \sigma ^z_{j+1} + \frac {h_x} 2(\sigma ^x_j  + \sigma^x_{j+1})+ \frac {h_z} 2 (\sigma^z_j + \sigma^z_{j+1})\;.
\end{align}
For $h_z = 0$  the model is free-fermion integrable (by Jordan-Wigner and Bogoliubov transforms);
the longitudinal field breaks that integrability.
Except where specified, I use $J = h_x = h_z = 1$.
These parameters give somewhat slower diffusion than the ``fruit-fly'' parameters of
\onlinecite{kim_ballistic_2013,kim_testing_2014,white_quantum_2018,rakovszky_dissipation-assisted_2020}.
Slower diffusion makes the small-system exact numerics cleaner, because the diffusive regime---after the short-time non-diffusive dynamics, but before the Thouless time---is larger.

%In Sec.~\ref{s:op-graph} 
% It is convenient to make a number of technical assumptions on the basis operators $\sigma^\alpha$,
% although my general results still hold (mutatis mutandis) if they are relaxed.
% I assume
% \begin{align}
%   \label{eq:basis-id}
%   I = \sigma^0\;.
% \end{align}
% I also assume that the $\sigma^\alpha$ have orthogonality and normalization properties
% \begin{align}
%   \tr [\sigma^{\alpha\dagger} \sigma^\beta] = d^{L} \delta_{\alpha\beta}
% \end{align}
% where $d$ is the dimension of the onsite Hilbert space,
% and that $[\sigma^\alpha, \sigma^\beta] \propto \sigma^\nu$ for at most one $\sigma^\nu$---that is, that the $\sigma^\alpha$ have structure constants
% \begin{align}
%   [\sigma^\alpha, \sigma^\beta] = if^{\alpha\beta\nu}\sigma^\nu\;,
% \end{align}
% such that for any particular $\alpha,\beta$, $f^{\alpha\beta\nu} \ne 0$ for at most one $\nu$.
% The qubit Pauli matrices satisfy these assumptions, as do the qudit Pauli matrices and the Gell-Mann matrices.
% The odd-onsite-dimension phase space point operators satisfy all these assumptions besides \eqref{eq:basis-id}.

The Ising Hamiltonian \eqref{eq:tfim} has only one conserved quantity, the energy density.
Since it is also translation invariant,
its infinite-temperature hydrodynamics is characterized by the dynamical correlation functions
\begin{align}
  \label{eq:eps-corr}
  \expct{\varepsilon_k(t) \varepsilon_k(0)} = \tr [\varepsilon_k(t) \varepsilon_k(0)]\;
\end{align}
where $\varepsilon_k = \sum_j \cos( k j) \varepsilon_j$, $k \in \{0, 2\pi/ L, \dots, \pi\}$ are Fourier modes of the energy density.
The diffusion equation \eqref{eq:diff} predicts
\begin{align}
  \varepsilon_k(t) = \varepsilon_0 e^{-Dk^2t}\;,
\end{align}
so one can measure the diffusion coefficient (or the presence of non-diffusive dynamics)
from the decay of the correlation functions \eqref{eq:eps-corr}.

\section{Liouvillian graph}\label{s:op-graph}

Correlation functions like \eqref{eq:eps-corr} are naturally phrased in terms of the Heisenberg dynamics of operators.
In this section I exactly map that Heisenberg dynamics to single-particle hopping on the ``Liouvillian graph'',
i.e. the graph whose vertices and edges are
\begin{align}
  V &= \{ \sigma^{\bm \mu} \equiv \sigma_1^{\mu_1} \dots \sigma^{\nu_L}_L \} \\
  E &= \{ (\bm \mu, \bm \nu) : \tr\left(\sigma^{\bm \nu} [H,\sigma^{\bm \mu}]\right) \ne 0\}\;.
\end{align}
I then show that that graph has a peculiar structure.
I organize the graph by string diameter:
the spatial extent of the region on which the Pauli string acts nontrivially.
The graph then consists of ``pools'' of strings, all with the same diameter, that---in the large-diameter limit---are much more thickly intraconnected than interconnected.
In particular the number of edges between a given operator $\sigma^{\bm \mu}$ of diameter $l$
operators of diameter $l+1$ and $l-1$ are
\begin{align}
  \begin{split}
    |\{\bm \nu  : (\bm \mu ,\bm \nu ) \in E, \diam \bm \nu  = l+1\}| &\le 12\\
    |\{\bm \nu  : (\bm \mu ,\bm \nu ) \in E, \diam \bm \nu  = l-1\}| &\le 6
  \end{split}
\end{align}
respectively, while the number of edges between $\sigma^{\bm \mu}$ and operators of length $l$ is
\begin{align}
  |\{\bm \nu : (\bm \mu, \bm \nu) \in E, \diam \bm \nu = l \}| \le 11l + 6\;.
\end{align}
(cf Fig.~\ref{fig:structure})
For the Ising model
\begin{align}
    n_+ &\equiv \overline{|\{\bm \nu  : (\bm \mu ,\bm \nu ) \in E, \diam \bm \nu  = l+1\}|} &&= \frac 4 3\\
    n_0 &\equiv \overline{|\{\bm \nu  : (\bm \mu ,\bm \nu ) \in E, \diam \bm \nu  = l\}|} &&= \frac 2 3 l\\
    n_- &\equiv \overline{|\{\bm \nu  : (\bm \mu ,\bm \nu ) \in E, \diam \bm \nu  = l-1\}|} &&= \frac 2 3\;,
\end{align}
where the overline $\overline{\cdot}$ indicates an average over strings of $\bm \mu$ of diameter $l$.

\subsection{Construction of the Liouvillian graph}

Consider the Heisenberg dynamics of an operator $A$.
Expand the operator $A$ in the basis of strings of operators
\begin{align}
  A &= \sum_{\bm \mu}  A_{\bm \mu}  \sigma ^{\bm \mu}
\end{align}
where the basis operators are $\sigma^{\bm \mu} = \sigma^{\mu_1} \dots \sigma^{\mu_L}$ and the coefficients are $A_{\bm \mu} = A_{\mu_1 \dots \mu_L}$.
Then
\begin{equation}
  \label{eq:heis}
  \frac d {dt} A_{\bm \mu}  = -i L_{\bm \mu \bm \nu }(t) A_{\bm \nu}
\end{equation}
where the $L_{\bm \mu \bm \nu}$ are the matrix elements of the Liouvillian superoperator
\begin{equation}
  \label{eq:liouv-mat-el}
 L_{\bm \mu \bm \nu }(t) = -2^{-L}\tr( \sigma ^{\bm \mu}  [H(t), \sigma^{\bm \nu} ]) \;.
\end{equation}
(The factor $2^{-L}$ is needed because $\tr [\sigma^{\bm \mu 2}] = 2^L$.)

One can understand this as single-particle hopping on a graph $G = (V,E)$ where the vertices are the Pauli indices $\bm \mu $ and the edge set is
\begin{equation}
  E = \{(\bm \nu \bm \mu ) : L_{\bm \nu \bm \mu } \ne 0\}\;.
\end{equation}
Following \onlinecite{cao_statistical_2021} I call this graph the \textbf{Liouvillian graph} of the Hamiltonian.
The Heisenberg dynamics \eqref{eq:heis}, \eqref{eq:liouv-mat-el} is then exactly equivalent to the dynamics of the single-particle Hamiltonian
\begin{equation}
  \label{eq:Hgraph}
  H_\graph(t) = \sum_{\langle \bm \mu \bm \nu  \rangle} L_{\bm \mu \bm \nu } c^\dagger_{\bm \mu}  c_{\bm \nu }\;.
\end{equation}
where $\langle \bm \mu \bm \nu  \rangle$ indicates that $\bm \mu $ and $\bm \nu $ are nearest neighbors in the graph $G$.

When discussing the dynamics of this Hamiltonian I will frequently write operators as states, e.g.
\begin{align}
  \label{eq:sigma-st}
  \ket{\bm \mu} = 2^{-L/2} \sigma^{\bm \nu}
\end{align}
and use the Frobenius inner product
\begin{align}
  \label{eq:frob}
  \braket{\bm \mu|\bm \nu} = 2^{-L} \tr[\sigma^{\bm \mu} \sigma^{\bm \nu}] = \delta_{\mu_1 \nu_1} \dots \delta_{\mu_L \nu_L}\;.
\end{align}
In general, as in Eq.~\eqref{eq:sigma-st},
I will take the states to be normalized with respect to the Frobenius inner product \eqref{eq:frob}.

\subsection{Structure of the Liouvillian graph} \label{ss:graph-structure}

To understand the system's dynamics, then, one needs to understand the structure of the Hamiltonian's Liouvillian graph.
The Liouvillian graph naturally decomposes into subgraphs of operators of fixed diameter
because an operator of diameter $l$ is connected to $O(l)$ other operators of length $l$, but $O(1)$ operators of diameter $l+1$ or $l-1$.
Call these subgraphs $\pool_l$.
%For $l \gg 1$, then, we should expect dynamics within each subgraph of diameter-$l$ operators to dominate dynamics connecting those subgraphs.

In fact the pool $\pool_l$ is itself the union of many disconnected subgraphs,
because a Pauli string with support in $\{j \dots j+l-1\}$
is connected to the Pauli strings with support in $\{j+1 \dots j+l\}$
only via strings of diameter $l-1$ or $l+1$.

To be precise, define the \textbf{support} of a Pauli string labeled by a multi-index $\bm \mu $ to be
\begin{equation}
  \supp \bm \mu  := \{j : \bm \mu _j \ne 0\}
\end{equation}
and its \textbf{diameter} to be
\begin{equation}
  \diam \bm \mu  := \max \supp \bm \mu  - \min \supp \bm \mu  + 1\;.
\end{equation}
Note that a diameter-$l$ string can have identities between its leftmost and rightmost nontrivial operators,
and may in fact only be nontrivial on those points. For example
\begin{equation}
  \label{eq:diam-id-ex}
  \sigma ^x_j\sigma ^x_{j+1}\quad\text{and}\quad \sigma ^x_j\sigma ^x_{j+17}
\end{equation}
have diameter 2 and 18 respectively.

Consider now a Pauli string $\sigma^{\bm \mu} $ with diameter $l$ and leftmost nontrivial site $j$---that is, $\supp \sigma^{\bm \mu} \subseteq j, \dots, j+l-1$ and $j, j+l-1 \in \supp \sigma^{\bm \mu}$.
It connects to operators
\begin{align}
  \label{eq:connect-to}
 \sigma^{\bm \nu} \propto [\sigma^\alpha_{k}\sigma^{\beta}_{k+1}, \sigma^{\bm \mu} ]\;.
\end{align}
for all $\alpha,\beta$ where $J^{\alpha\beta}_k \ne 0$.
Divide the set of such operators $\sigma^{\bm \nu}$ up by diameter,
defining notation
\begin{align}
  \begin{split}
    \mathcal M_{l'}(\bm \mu)
    &= \{
    \bm \nu : \text{$\diam \bm \nu = l'$ and $L_{\bm \nu\bm \mu} \ne 0$}
    \}
  \end{split}
\end{align}
for the set of length-$l'$ operators to which $\bm \mu$ is connected by the Liouvillian.
Now the set of all operators $\sigma^{\bm \nu}$ of \eqref{eq:connect-to} is
\begin{align}
  \{ \bm \nu : L_{\bm \mu \bm \nu} \ne 0\} = \mathcal M_{l-1}(\bm\mu) \cup \mathcal M_{l}(\bm\mu) \cup \mathcal M_{l+1}(\bm\mu)\;.
\end{align}
A particular $\sigma^{\bm \nu}$ is $\sigma^{\bm \nu} \in \mathcal M_{l+1}$,
i.e. has diameter $l+1$, only if
$k = j-1$ and $\alpha \ne 0$, $\beta \not\in \{ 0,\mu_j\}$
(or $k = j + l - 1$, $\alpha \not\in\{0,\mu_{j+l-1}\}$, $\beta \ne 0$).
Similarly
$\sigma^{\bm \nu}$ has diameter $l-1$ only if
$k = j$ and $\alpha = \mu_j, \beta \not\in\{0,\mu_{j+1}\}$
(or $k = j+l-2$ and $\alpha \not\in\{0,\mu_{j+l-2}\}, \beta = \mu_{j+l-1}$).
The starting operator $\sigma^{\bm \mu}$ therefore connects to 
\begin{align}
  \label{eq:interpool-bound}
  \begin{split}
    |\mathcal M_{l+1}(\bm \mu)| &\le 12\\
    |\mathcal M_{l-1}(\bm \mu)| &\le 6
    %|\{\bm \nu  : (\bm \mu ,\bm \nu ) \in E, \diam \bm \nu  = l-1\}| &\le 6
  \end{split}
\end{align}
operators of length $l+1$, $l-1$ respectively.
(Since many of the coefficients $J^{(k)}_{\alpha\beta}$ of these terms $\sigma^{\alpha}_k\sigma^{\beta}_k$ will be zero for physical Hamiltonians, these bounds will not be saturated.
At the same time we can loosely bound
\begin{align}
  \label{eq:intrapool-bound}
    |\mathcal M_{l}(\bm \mu)| &\le 11 l + 9\;,
\end{align}
accounting separately for single-site and two-site operators in $J_{\alpha\beta}$.

Most strings have some number of identity operators between their left- and right-most nontrivial operators,
so they will not saturate the bound \eqref{eq:intrapool-bound}.
The string $\sigma^x_j\sigma^x_{j+17}$ of Eq.~\eqref{eq:diam-id-ex} is an extreme example of this situation.
But strings like $\sigma^x_j\sigma^x_{j+17}$ are atypical.
The average number of nontrivial Pauli operators in a randomly chosen Pauli string of diameter $l$ will be
\[ \frac 3 4 l - \frac 1 2\;, \] %2 + 3/4 (l - 2)
so we expect an average number of intrapool connections
\begin{align}
  \label{eq:intrapool-degree}
  n_0 \equiv \overline{|\mathcal M_{l}|} = a l + b\
\end{align}
for some constants $a,b$.
The overline indicates an average over starting strings $\bm \mu$.

\begin{figure}[t!]

  \begin{minipage}{0.45\textwidth}
    \includegraphics[width=\textwidth]{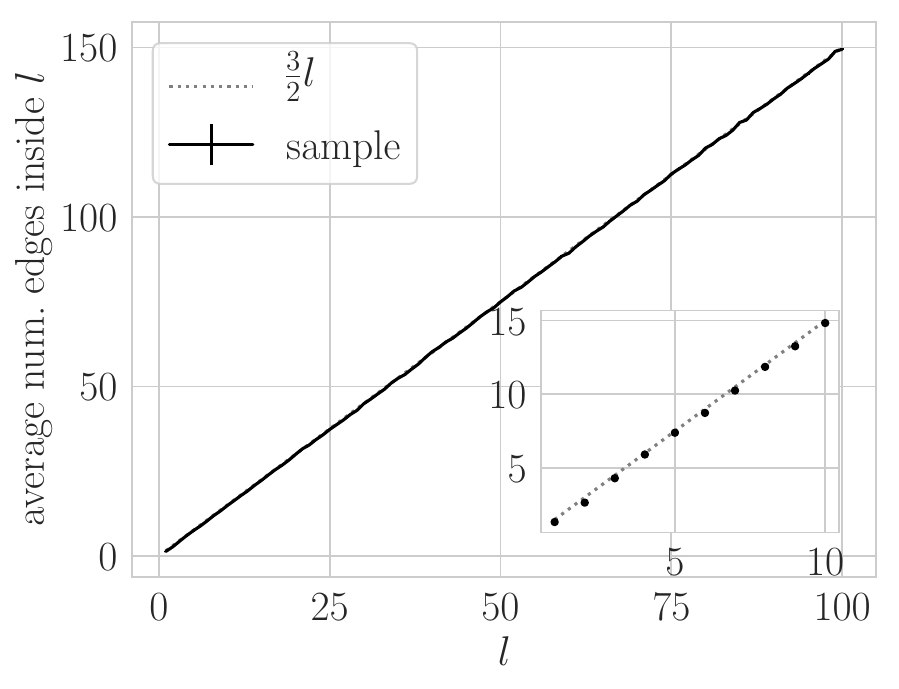}
  \end{minipage}

  \begin{minipage}{0.45\textwidth}
    \includegraphics[width=\textwidth]{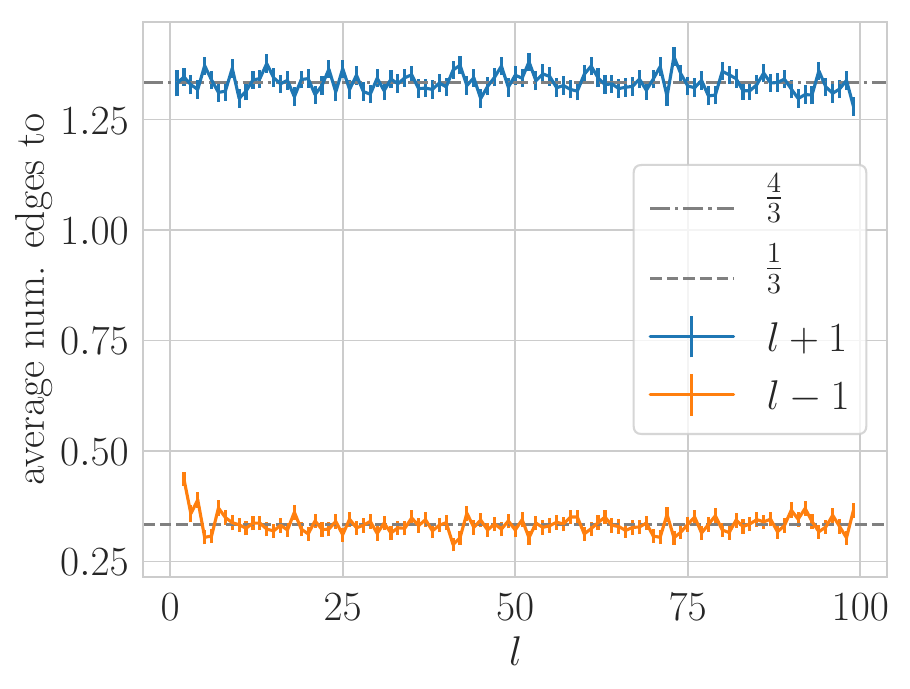}
  \end{minipage}
  
  \caption{\textbf{Distribution of edges} in the Liouvillian graph of the transverse-field Ising model \eqref{eq:tfim}.
    \textbf{Top:} average number of edges connecting a length-$l$ Pauli string to other length-$l$ Pauli strings.
    \textbf{Bottom:} average number of edges connecting a length-$l$ operator to a length-$l+1$ operator (blue) or a length-$l-1$ operator (orange).
    For $l \gg 1$, an operator has many more connections to other length-$l$ operators than to operators of other lengths.
  }
  \label{fig:internal-connections}
\end{figure}

By reasoning like that leading to Eq.~\eqref{eq:interpool-bound} one can show that for the Ising model \eqref{eq:tfim}
the average number of connections from pool $\pool_l$ to $\pool_{l-1}$, $\pool_{l+1}$, and $\pool_{l}$  are
\begin{align}
  n_- = \frac 1 3 \qquad n_+ = \frac 4 3  \qquad n_0 = \frac 3 2 l\;.
\end{align}
respectively.
In Fig.~\ref{fig:internal-connections} I show a numerical check.
I estimate $n_\pm, n_0$ by sampling
which lets me work with large systems.
(I take $L = 100$, but more would not be unreasonable.)
Fig.~\ref{fig:internal-connections} top shows $n_0$ in black;
Fig.~\ref{fig:internal-connections} bottom shows $n_+$ in blue and $n_-$ in orange.

\section{Operator weight escape rate} \label{s:rate-loss}

We ultimately seek the rate at which operator weight escapes into long operators.
In the language of the Liouvillian graph Hamiltonian,
this is the rate at which the single particle escapes from the pool $\pool_{l-1}$ of Pauli strings of length $l-1$ into the pool $\pool_l$ of Pauli strings of length $l$.

In this section I argue that that rate is controlled by
the characteristic timescale of the coherent dynamics of the pool $\pool_l$
via a quantum-Zeno-like effect.
I first illustrate that effect with a toy model.
I then estimate the characteristic timescale for dynamics in $\pool_l$,
assuming that that dynamics is chaotic in a certain sense,
and check that prediction against numerical calculations in the Ising model \eqref{eq:tfim}.
Finally I make a somewhat more careful calculation of the escape rate,
and again compare to numerics.

These calculations make no assumptions about ``backflow''---the return of operator weight from long operators to short operators.
Rather, I estimate the rate at which operator weight escapes into long operators in the first place.

\subsection{Toy model}\label{ss:loss-toy}

Imagine a single particle on two sites $a,b$ connected by a hopping amplitude---that is, governed by a Hamiltonian
\begin{align} 
  H =  V c^\dagger_a c_b + h.c.\;.
\end{align}
Suppose the particle is lost at some rate $\gamma_b$ from site $b$,
in the same way that a particle on the Liouvillian graph can hop from a site in $\pool_l$ to one in $\pool_{l+1}$ and then disappear into the rest of $\pool_{l+1}$.
Mock this up by endowing site $b$ with a self-energy
\begin{align}
  S_b(\omega) = -i\gamma\;.
\end{align}
(In principle this self-energy depends on frequency $\omega$;
here I consider only the leading long-time $\omega \to 0$ behavior.)
Follow (very broadly) the logic of \onlinecite{abou-chacra_selfconsistent,feenberg_note_1948,watson_multiple_1957}.
At leading order in $V$ (that is, for $V \ll \gamma$) the self-energy of $a$ is
\begin{align}
  S_a(\omega) \approx \frac {|V|^2}{\omega - i\gamma} \;.
\end{align}
If the particle starts on site $a$, then,
Site $b$ induces an effective loss rate
\begin{align}
  \label{eq:toy-loss-rate}
  S_a(\omega = 0) = i\gamma^{-1}|V|^2\;:
\end{align}
if the particle disappears quickly from site $b$, it will disappear slowly from site $a$, and vice versa.
(I give a more detailed version of this argument in App.~\ref{app:toy}).

\subsection{Estimating intrapool decay rate}\label{ss:rate-loss-est}

\begin{figure}
  \begin{minipage}{0.45\textwidth}
    \includegraphics[width=\textwidth]{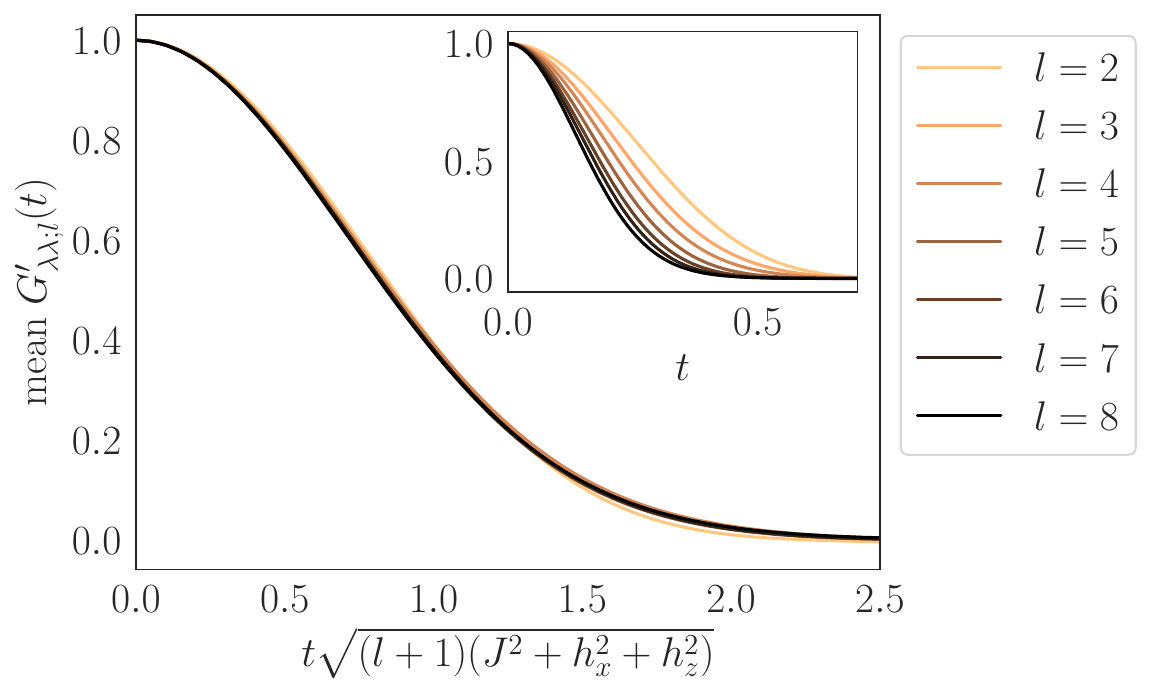}
  \end{minipage}
  
  \begin{minipage}{0.45\textwidth}
    \includegraphics[width=\textwidth]{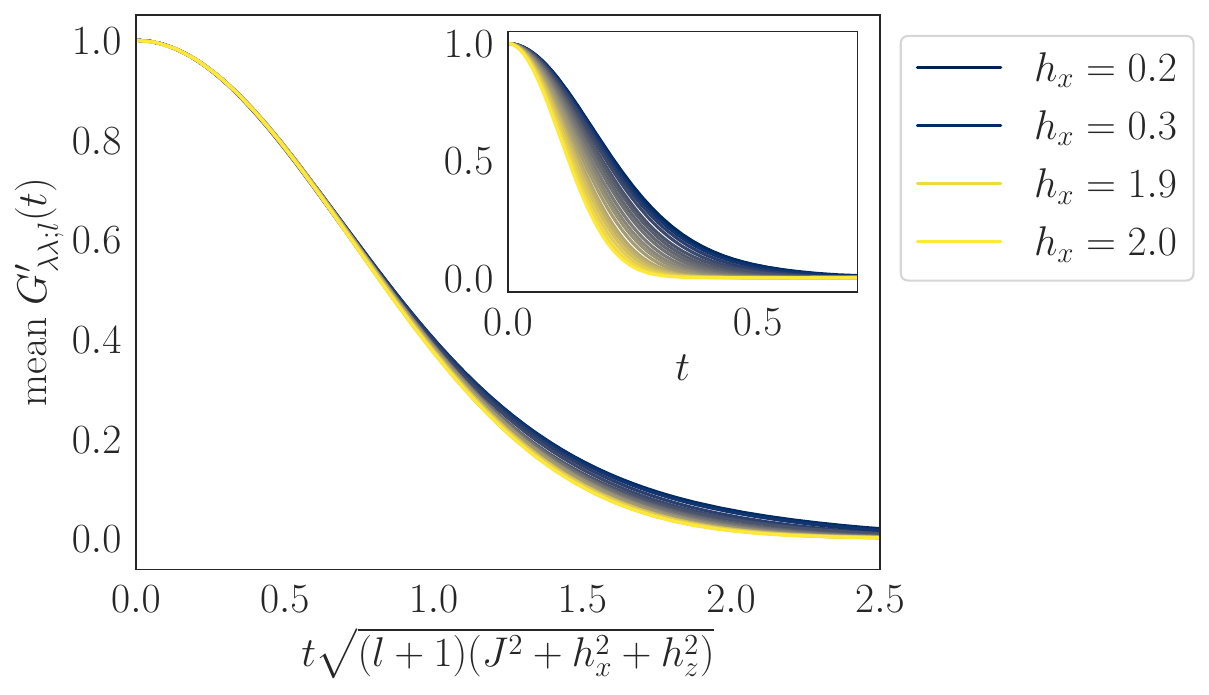}
  \end{minipage}

  \caption{\textbf{Single-site average return amplitude} $ G'_{\bm\lambda\bm\lambda;l}(t)$
    (Eq.~\eqref{eq:single-site-ret-prime})
    for dynamics restricted to the subgraph of operators with diameter $l,l+1$,
    as a function of time and operator length,
    for transverse field $h_x = 1$ (\textbf{top}),
    and transverse field for operator length $l = 8$ (\textbf{bottom}).
    In each case I take $h_z = 1.0$.
    For $l \le 4$ I average over all initial operators $\ket{\bm \mu}$;
    for $l > 4$ I average over $512$ randomly chosen initial operators.
    In each case the inset plots the amplitude against time \textit{simpliciter},
    while the main axes plot it against time rescaled by the semi-empirical prediction of Eq.~\eqref{eq:semi-emp-2}.
    The return amplitude is well-characterized by the single timescale $\gamma_{\text{semi-emp}}^{-1}$.
  }
  \label{fig:self-energy-average}
\end{figure}

We therefore require the rate at which a particle spreads from a site $\bm \lambda \in \pool_{l}$
to the other operators in $\pool_l$.
To characterize this decay rate, consider the amplitude for the particle to remain on the initial state.
It is easiest to proceed in real time, rather than Fourier space;
in that language we require
\begin{align}
  \label{eq:ret-ampl-defn}
  G_{\bm\lambda\bm\lambda;l}(t) = \bra{\bm \lambda} e^{-iH_{\graph;l}t}\ket{\bm \lambda}\;,
\end{align}
where I write $H_{\graph;l}$ for the graph Hamiltonian restricted to the subgraph $\pool_l$.

Assume that $H_{\graph;l}$ is chaotic, in the sense that it has extended eigenstates
\begin{align}
  \label{eq:chaos}
  |\braket {\bm \lambda| E_\alpha}|^2 \approx \frac 1 {| \pool_l|}
\end{align}
for all $\lambda \in \pool_l$.
One expects this assumption to be better for large $l$.
Then
\begin{align}
  \label{eq:ret-ampl}
  G_{\bm\lambda\bm\lambda;l}(t) = \int dE\; \rho(E) e^{-iH_{\graph;l} t} \;.
\end{align}
(This continuum approximation, like the chaos assumption, is better justified for larger $l$.)
By the uncertainty principle for Fourier transforms the decay rate will be given by the width of the density of states
\begin{align}
  \gamma^2 \propto \int dE E^2 \rho(E) = \frac 1 {|\pool_l|} \tr H_{\graph;l}^2 \;.
\end{align}
But
\begin{subequations}
  \begin{align}
    \tr H_{\graph;l} &= 0 \\
    \label{eq:trH2}
    \begin{split}
     \Delta^2 \equiv \frac 1 {|\pool_l|} \tr H_{\graph;l}^2 &= \sum_{\bm \mu, \bm \nu \in \pool_l} L_{\bm\mu\bm\nu}^2\\
                     &\approx n_0 \bar J^2 \\
                     &= (al + b) \bar J^2\;,
                   \end{split}
  \end{align}
\end{subequations}
where { I define notation $\Delta$ for the standard deviation of the density of states,
$n_0 = al + b$ is the average degree of $\pool_l$ (cf Eq.~\eqref{eq:intrapool-degree}),
and $\bar J$ is the mean in quadrature of the nonzero terms in the Hamiltonian \eqref{eq:nn-ham}
\begin{equation}
  \bar J^2 = \frac 1 {|\{ J_{\alpha\beta} \ne 0\}|} \sum_{J_{\alpha\beta} \ne 0} J_{\alpha\beta}^2\;.
\end{equation}
For notational simplicity I here take the Hamiltonian to be translation invariant.
Thus $ \gamma \propto \sqrt{al + b} \bar J$.

To be more precise, assume that $H_{\graph;l}$ has a Gaussian density of states
\begin{align}
  \label{eq:gaussian-dos}
  \rho(E) = \frac 1 {\Delta \sqrt{2\pi}} e^{-E^2/2\Delta^2}\;.
\end{align}
$\Delta$ is given by \eqref{eq:trH2},
and the return amplitude of \eqref{eq:ret-ampl} is
\begin{align}
  \label{eq:gaussian-ret-ampl}
  G_{\bm\lambda\bm\lambda;l}(t) = e^{-\gamma^2 t^2}
\end{align}
with decay rate
\begin{align}
  \gamma^2 = 2(al + b) \bar J^2\;.
\end{align}
For the transverse-field Ising model 
\begin{align}
  a = \frac 3 2,\qquad b = 0
\end{align}
and
\begin{align}
  \bar J = \frac 1 3 \sqrt{1 + h_x^2 + h_z^2}\;.
\end{align}
for
\begin{align}
  \label{eq:ising-gamma-pred}
  \gamma = \sqrt{l(1 + h_x^2 + h_z^2)}\;.
\end{align}

\begin{figure}[t]
  % \begin{minipage}{0.45\textwidth}
  %   \includegraphics[width=\textwidth]{ODETRUNC019_fit-gamma-l.pdf}
  % \end{minipage}

  \begin{minipage}{0.45\textwidth}
    \includegraphics[width=\textwidth]{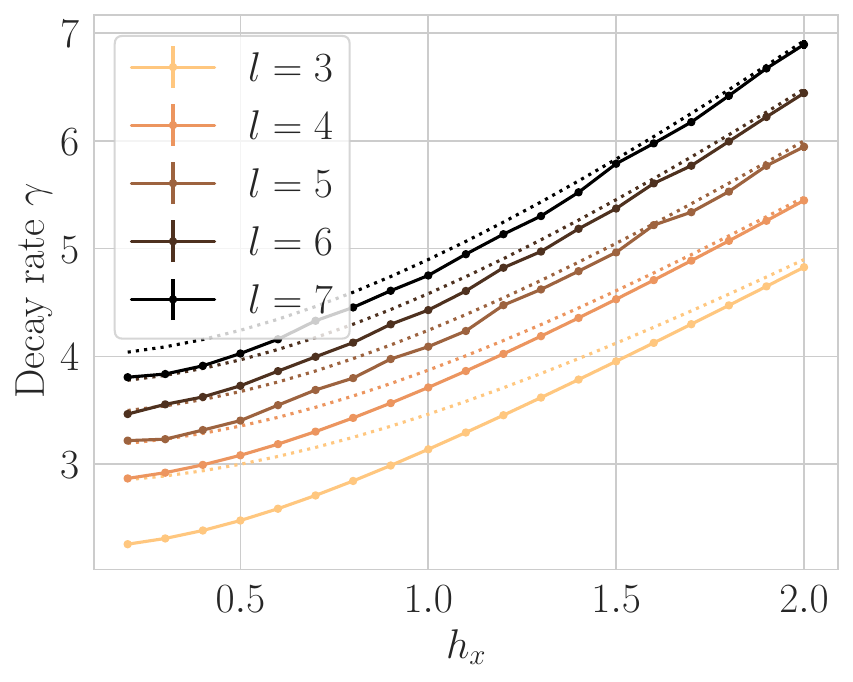}
  \end{minipage}

  \caption{\textbf{Decay rates extracted from Gaussian fits} to the numerical average $G'_{\bm\lambda\bm\lambda;l}(t)$ of Fig.~\ref{fig:self-energy-average}, as a function of operator length $l$ and transverse field $h_x$.
    For $l\gg 1$ these fits agree well with the semi-empirical decay rate $\gamma_{\text{semi-emp}}^{-1}$ of Eq.~\eqref{eq:semi-emp-2} (dotted line).
  }
  \label{fig:decay-fit}
\end{figure}

Figures \ref{fig:self-energy-average} and \ref{fig:decay-fit} show numerical checks of this prediction for the transverse-field Ising model.
I seek to verify the predictions \eqref{eq:gaussian-ret-ampl}, \eqref{eq:ising-gamma-pred},
but also to validate that they accurately reflect the dynamics of the chain.
In App.~\ref{ss:fact-loss} I argue that beyond the shortest times, particle escape from $\pool_l$ to $\pool_{l+1}$
has an important effect on particle escape from $\pool_{l-1}$ to $\pool_l$.
I therefore consider the dynamics not only of $\pool_l$,
but also of $\pool_{l+1}$: that is, the return amplitude
\begin{align}
  \label{eq:single-site-ret-prime}
  G_{\bm\lambda\bm\lambda;l}'(t) := \bra{\bm \lambda} e^{-iH'_{\graph;l}t} \ket{\bm \lambda}
\end{align}
for $\bm \lambda \in \pool_l$, where $H'_{\graph;l}$ is the graph Hamiltonian restricted to $\pool_l \cup \pool_{l+1}$
\begin{align}
  H'_{\graph;l} = H_\graph |_{\pool_l \cup \pool_{l+1}}\;.
\end{align}

Fig.~\ref{fig:self-energy-average} shows
the average of the return amplitude $G'_{\bm \lambda \bm \lambda; l}(t)$
across starting operators $\bm \lambda$
as a function of time, for a variety of $l$ and $h_x$.
The prediction \eqref{eq:ising-gamma-pred} does not produce a good scaling collapse.
Instead, I rescale by
\begin{align}
  \label{eq:semi-emp-2}
  \gamma_{\semiemp} = \sqrt{(l+1)(J + h_x^2 + h_z^2)}\;.
\end{align}
The dynamics is well-characterized by this single timescale.

Fig.~\ref{fig:decay-fit} shows the decay rates $\gamma$ resulting from fitting the Gaussian \eqref{eq:gaussian-ret-ampl} to the numerical average return amplitude.
For $l \gg 1$ and $h_x \gtrsim 1$ I find that the fits agree well with
the decay rate $\gamma_{\semiemp}$ of \eqref{eq:semi-emp-2}.

In both Figures \ref{fig:self-energy-average} and \ref{fig:decay-fit} we see
\begin{align}
  \gamma \propto \sqrt{l+1}
\end{align}
rather than the predicted
\begin{align}
  \gamma \propto \sqrt{l}\;.
\end{align}
To understand this, return to the early-time argument of Sec.~\ref{ss:fact-loss}.
The scaling in operator diameter $l$ comes from the degree of the starting operator as a vertex in the Liouvillian graph.
For times $t \ll \gamma$, the state is
\begin{align}
  \ket \psi = \ket{\bm\mu} - it \sum_{\bm \nu} L_{\bm\nu\bm\mu}\ket{\bm \nu}\;.
\end{align}
But the full dynamics includes not only hopping from $\bm \mu$ to other operators $\bm \nu \in \pool_l$,
but also to operators in $\pool_{l+1}$.
One therefore expects the rate to go as
\begin{align}
  \gamma^2 \propto | \{\bm \nu : \bra {\bm \nu} H_\graph \ket{\bm \mu} \ne 0 \}|\;,
\end{align}
including $\ket{\bm \nu} \in \pool_{l+1}$:
that is, we should understand the degree as including not only the $O(l)$ connections to other operators in $\pool_l$,
but also the $O(1)$ connections to other operators in $\pool_{l+1}$, for
\begin{align}
  \gamma \propto \sqrt{l+a}
\end{align}
for some constant $c$.

\begin{figure}[t]
  \includegraphics[width=0.45\textwidth]{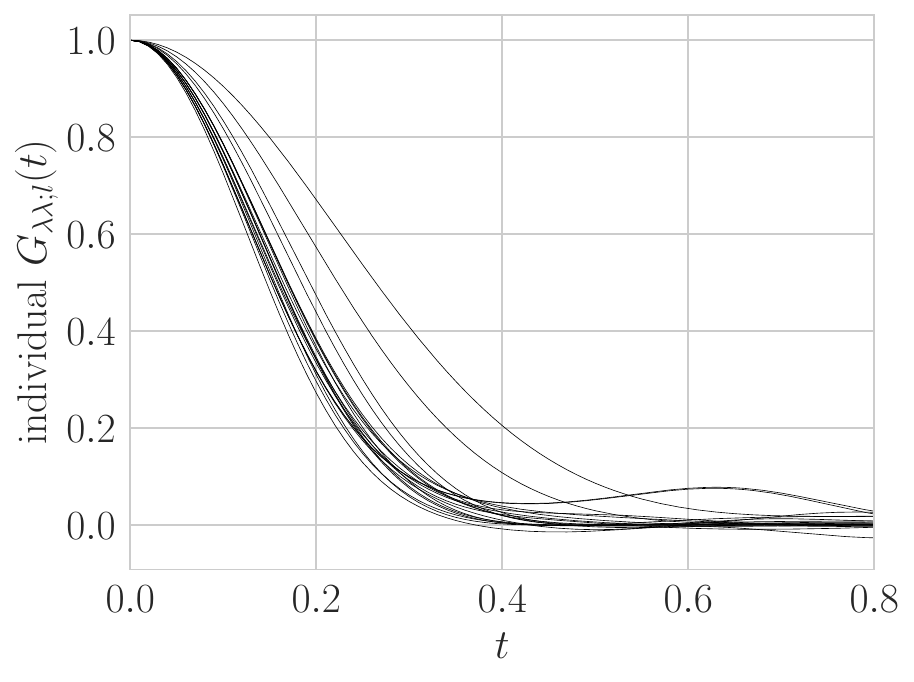}
  \caption{\textbf{Propagator $G'_{\bm \lambda\bm\lambda;l}(t)$}  (Eq.~\eqref{eq:single-site-ret-prime}) for 20 randomly chosen initial operators of diameter $l = 8$
     at $h_x = h_z = 1$.
    All show generally Gaussian decay, with consistent decay rate. }
  \label{fig:self-energy-individual}
\end{figure}

Averaging over starting operators as in Fig.~\ref{fig:self-energy-average} produces a clean characterization of the decay,
at the cost of concealing deviations from the chaos assumption \eqref{eq:chaos}:
in fact
\begin{equation}
  \label{eq:avg-trace}
  \sum_{\bm \mu \in \pool_l \cup \pool_{l+1} } G_{\bm \mu\bm \mu;l}'(t) = \tr e^{-iH'_{\graph;l} t}\;.
\end{equation}
Since I do not allow the starting string $\ket{\bm\mu}$ to reside in $\pool_{l+1}$
the measurements of Fig.~\ref{fig:self-energy-average} do not strictly speaking give the trace of Eq.~\eqref{eq:avg-trace},
but the same intuition applies.
  
In Fig.~\ref{fig:self-energy-individual}
I therefore plot individual propagators $G_{\bm\lambda\bm\lambda;l}'$
for 20 randomly chosen initial operators $\ket{\bm \lambda}$ with diameter $l = 8$.
All display a generally Gaussian decay, consistent with Eq.~\eqref{eq:gaussian-ret-ampl},
and all display generally the same decay rate.
This indicates that---at least for $l=8$---the chaos assumption \eqref{eq:chaos} is reasonable,
and the averages of Fig.~\ref{fig:self-energy-average}
(not to mention the single decay rate $\gamma_\semiemp$)
give an accurate characterization of the intrapool decay.

% \subsection{Basis independence}\label{ss:basis-dep}
% The operator graph clearly depends on the basis we chose for the onsite operators,
% and so (e.g.) on the basis we choose for the onsite state Hilbert space.
% Consider, for example, the unitary
% \begin{align}
%   U = \prod_j \frac 1 \sqrt{2} (1 + i\sigma^z_j)\;.
% \end{align}
% Under this unitary the transverse field Ising Hamiltonian of Eq.~\eqref{eq:tfim} becomes
% \begin{align}
%   H = \sum_j \sigma^z_j \sigma^z_j
%   + h_z \sum_j \sigma^z_j
%   + \frac {h_x}{\sqrt 2} \sum_j \sigma^x_j
%   + \frac {h_x}{\sqrt 2} \sum_j \sigma^y_j\;.
% \end{align}
% \todo{gah check}
% This will increase the degree of the operator graph.
% quantities of interest, e.g. $\tr H_{\graph;l}^2$, invariant under (superoperator) unitaries.

\subsection{Estimating the interpool escape rate}

\begin{figure}
  \begin{minipage}{0.45\textwidth}
    \includegraphics[width=\textwidth]{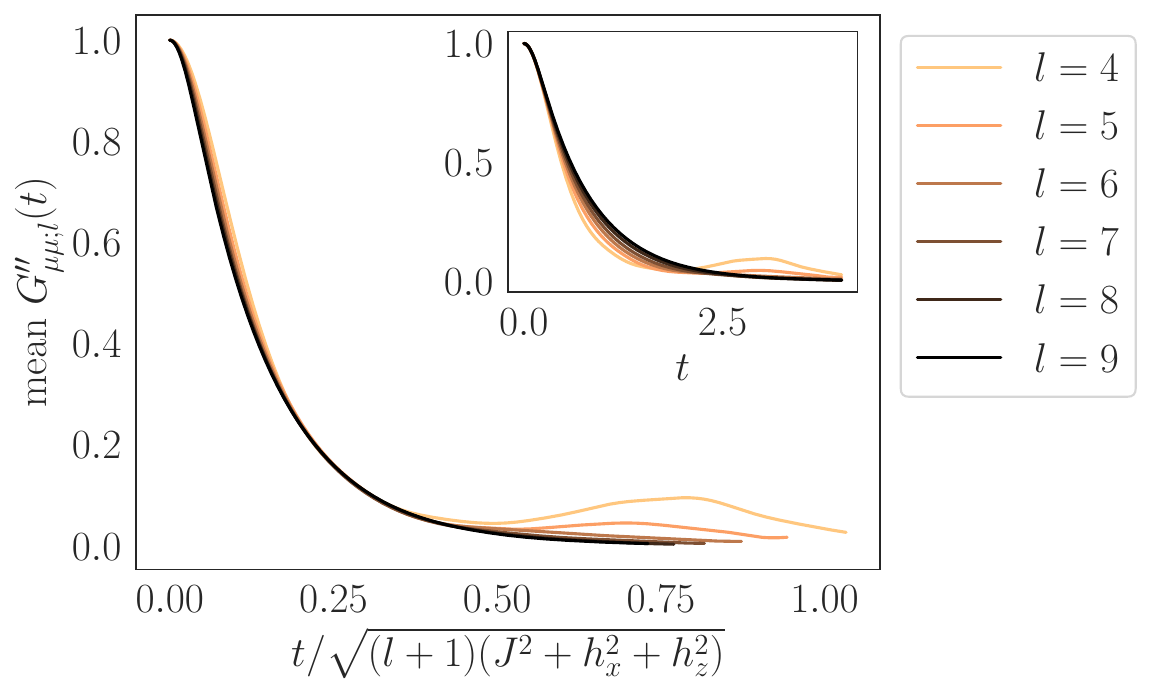}
  \end{minipage}

  \begin{minipage}{0.45\textwidth}
    \includegraphics[width=\textwidth]{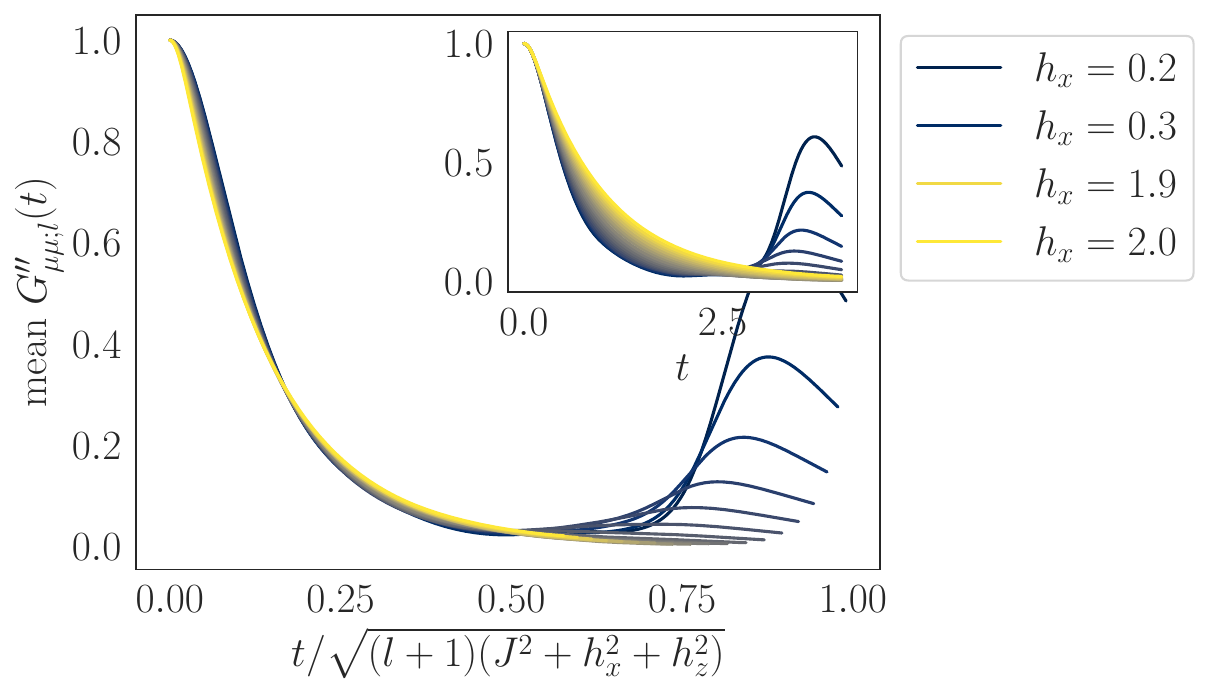}
  \end{minipage}
  \caption{\textbf{Escape from pool of diameter $l-1$}:
    onsite propagator $G''_{\mu\mu}$ through operators in pool of diameter $l$ (Eq.~\eqref{eq:escape-prop})
    as a function of diameter (\textbf{top}) and transverse field $h_x$ (\textbf{bottom}).
    For $l \le 5$ I average over all initial operators $\ket{\bm \mu}$;
    for $l > 5$ I average over $512$ randomly chosen initial operators. 
    I average over starting strings $\ket {\bm \mu} \in \pool_{l-1}$,
    chosen to connect on both ends to $\pool_{l}$.
    In each case rescaling by $\gamma_\semiemp$ as in \eqref{eq:esc-est} produces good results for intermediate times.
  }
  \label{fig:escape}
\end{figure}

I can now return to the question of how quickly a particle escapes from pool $\pool_{l-1}$ into $\pool_l$.
Take some $\bm \mu \in \pool_{l-1}$;
$\bm \mu$ has matrix elements $L_{\bm\mu\bm\lambda}$ with Pauli strings $\bm \lambda \in \pool_l$.
These come from ``bond terms'' in the Hamiltonian;
write $V$ for their characteristic scale.
Consider the part of the propagator $\bra {\bm \mu} e^{-iH_\graph t} \ket{\bm \mu}$ that passes through pools $\pool_l, \pool_{l+1}$:
that is, write a restricted Hamiltonian
\begin{align}
  H_{\graph;\bm \mu,l}'' = H|_{\{\bm \mu\} \cup \pool_{l} \cup \pool_{l+1} }
\end{align}
and compute the propagator
\begin{align}
  \label{eq:escape-prop}
  G''_{\bm \mu \bm \mu; l}(t) = \bra {\bm \mu} e^{-iH''_{\graph;\bm \mu,l} t} \ket{\bm \mu}\;.
\end{align}
At leading order in $Vt$ this is given by the irreducible part 
\begin{align}
  \begin{split}
    G''_{\bm \mu \bm \mu; l}(t) 
  %&\bra {\bm \mu} e^{-iH_{\graph;\bm \mu,l} t} \ket{\bm \mu}\\
  &\approx 1 
  - n_+%\sum_{\substack{\langle \bm \mu \bm \lambda\rangle, \\\bm \lambda \in \pool_l}}
  \int_0^t dt_2 \int_0^{t_1} dt_1\; V^2 e^{-\gamma^2(t_2 - t_1)^2}\\
  &\qquad + O(V^4t^4)\;.
\end{split}\\
  \label{eq:esc}
  &\approx 1 - 
  \begin{cases}
    n_+t^2V^2 & t\gamma \ll 1 \\
    t\,\gamma^{-1}  n_+ V^2 \sqrt{\pi/2} & t\gamma \gg 1
    \end{cases}
\end{align}
---that is, for $t\gamma \gg 1$ the escape rate from $\pool_l$ to $\pool_{l+1}$ is
\begin{equation}
  \label{eq:esc-est}
  \propto n_+V^2 \gamma^{-1}\;.
\end{equation}
similarly to the toy model (Eq.~\eqref{eq:toy-loss-rate}).

I plot the average of $G''_{\bm\mu\bm\mu;l}(t)$ across starting operators $\ket {\bm \mu}$ in Fig.~\ref{fig:escape}.
I choose starting strings that begin and end with $\sigma^x$ or $\sigma^y$,
so all operators connect to the pool $\pool_{l}$ on both ends.
For short times all the different parameter values ($h_x, l$) display the same quadratic behavior,
and for intermediate times all collapse when rescaled \`a la \eqref{eq:esc-est}.
Both limits are therefore consistent with \eqref{eq:esc}.
For longer times, $l \gtrsim 6$ and $h_x \gtrsim 0.6$ continue to display a good scaling collapse,
but $l \lesssim 5$ and $h_x \lesssim 0.5$  do not.
The small-$h_x$ behavior comes about because the $h_x$ term is the only term that fails to commute with $\sigma^z$:
in the limit $h_x \to 0$ amplitude starting on a Pauli string $\ket{\bm \mu} = \ket{\dots \sigma^x\sigma^0\dots}\in\pool_{l-1}$ will oscillate between that operator and $\ket{\bm \lambda} = \ket{\dots \sigma^y\sigma^z}$.
I expect that the small-$l$ behavior comes about because both the chaos assumption \eqref{eq:chaos} and the continuum density of states approximation of \eqref{eq:ret-ampl} break down when the pool $\pool_l$ has few operators.

\begin{figure}[t]
  \centering
  \includegraphics[width=0.45\textwidth]{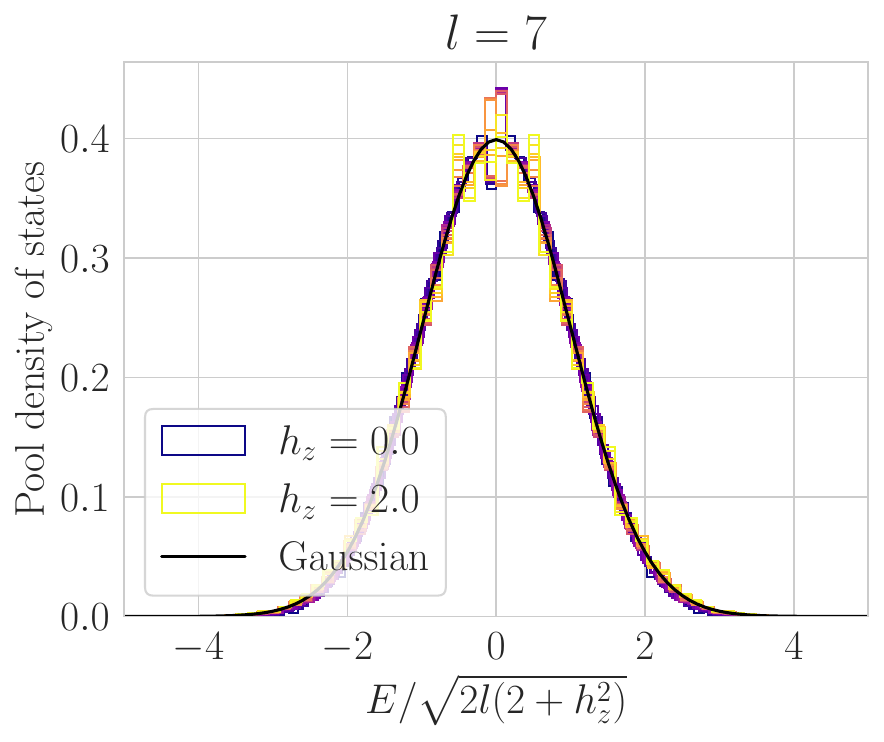}
  \caption{\textbf{Density of states} for the $l = 7$ pool Hamiltonian in the model \eqref{eq:tfim} for $h_x = 1$ and $h_z = 0.0, 0.1, \dots, 1.9, 2.0$. Energies are rescaled by the standard deviation of Eq.~\eqref{eq:trH2}; black line shows a standard Gaussian.}
  \label{fig:gaussian-dos}
\end{figure}

\subsection{ Breakdown of chaos assumptions}\label{ss:chaos-breakdown}

\begin{figure}[t]
  \centering
  \begin{minipage}{0.45\textwidth} 
    \includegraphics[width=\textwidth]{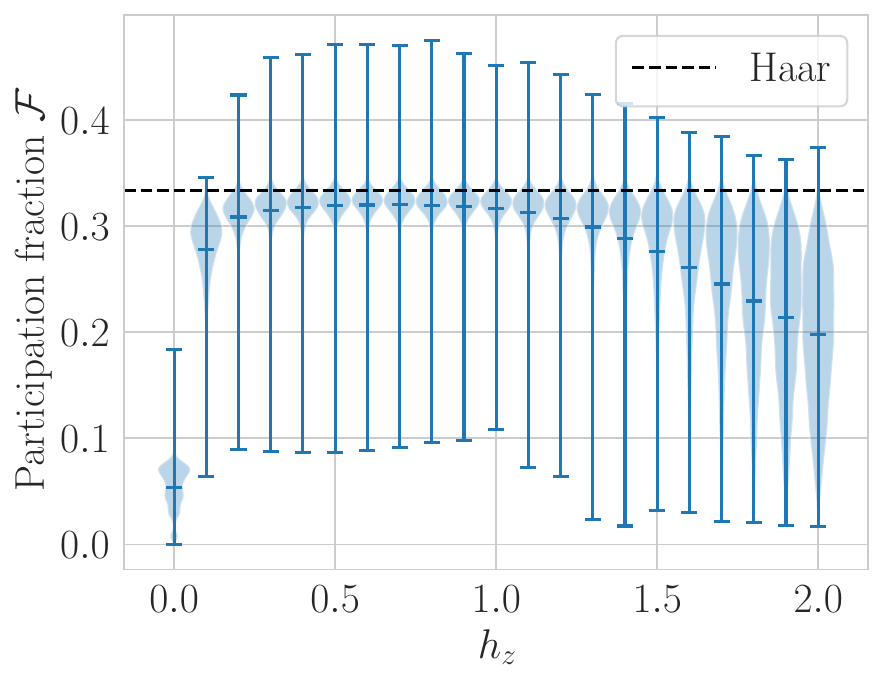}
  \end{minipage}
  \begin{minipage}{0.45\textwidth}
  \includegraphics[width=\textwidth]{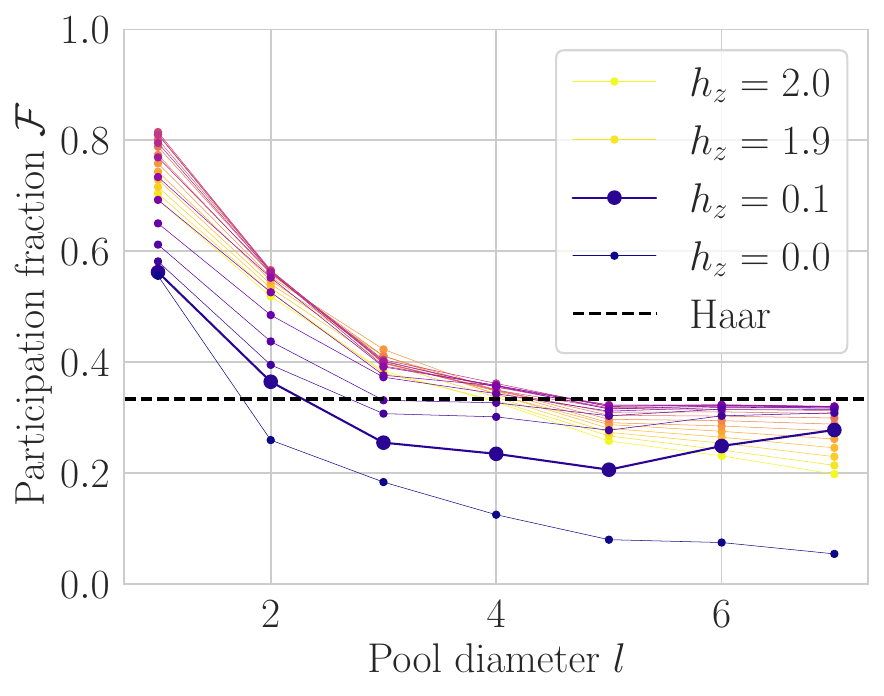}
  \end{minipage}
  
  \caption{
    \textbf{Pool eigenstate participation fraction}.
    \textbf{Top:} Violin plot of the distribution of pool eigenstate Hilbert space fractions \eqref{eq:H-frac}
    as a function of longitudinal field $h_z$ (blue) at pool diameter $l = 7$.
    Horizontal lines in the middle of the distribution show means;
    caps show outliers, not standard deviation.
    The pool Hamiltonian at this diameter is chaotic for $0.25 \lesssim h_z \lesssim 1.25$.
    \textbf{Bottom:} mean participation fraction as a function of $l$, across $h_z$.
    The bold line, $h_z = 0.1$, displays non-monotonicity resulting from weakly-broken free fermion integrability.
    }
  \label{fig:participation-fraction}
\end{figure}

\begin{figure}
  \includegraphics[width=0.45\textwidth]{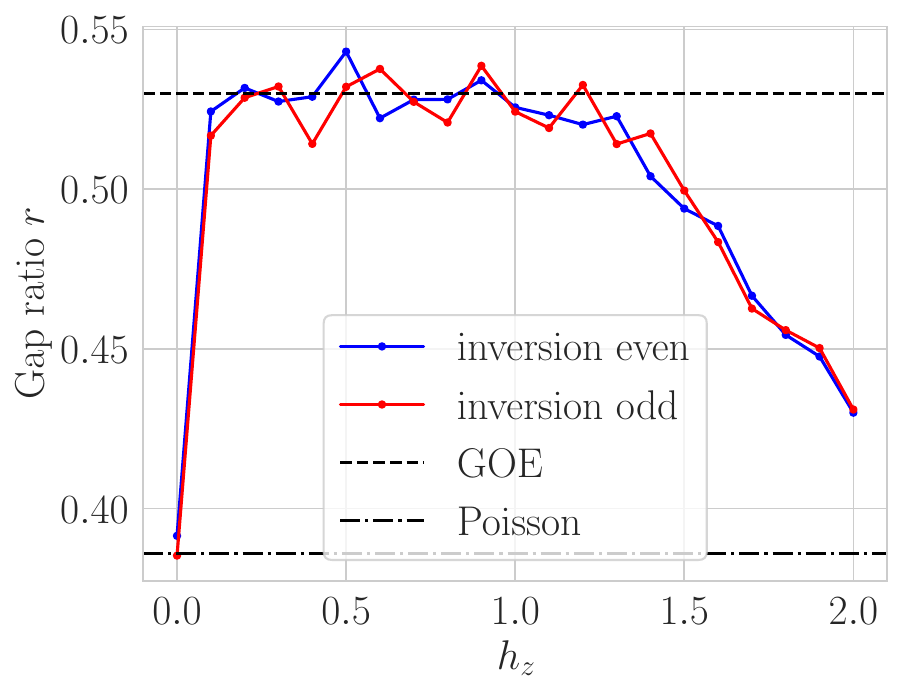}
  \caption{
    \textbf{Pool Hamiltonian eigenenergy gap ratio} for the pool of diameter $l = 7$ as a function of $h_z$.
    The system is inversion symmetric, so I plot the inversion-even (blue) and inversion-odd (red) sectors separately.
    Again the pool Hamiltonian at this diameter is chaotic for $0.25 \lesssim h_z \lesssim 1.25$.
  }
  \label{fig:clean-gap-ratio}
\end{figure}

  Sec.~\ref{ss:rate-loss-est} used two assumptions about the dynamics of long pools:
  that the density of states is Gaussian (Eq.~\eqref{eq:gaussian-dos}),
  and that the pool eigenstates are extended (Eq.~\eqref{eq:chaos})%
  ---that is, that the pool Hamiltonians are chaotic.

  The Gaussian shape of the density of shapes is a technicality.
  If the density of states has a different shape,
  the decay rate differs by some dimensionless factor,
  but it depends on the pool diameter and the Hamiltonian parameters in broadly the same way.

  The chaos assumption is not a technicality.
  If it is not satisfied, the pool eigenstates must have some structure.
  In that case the pool dynamics are controlled by that structure,
  rather than by the single decay rate $\gamma^2 \propto \tr H_{\graph;l}^2$.
  
  In this section I numerically check these assumptions by full diagonalization of the pool Hamiltonians.
  The density of states is always (approximately) Gaussian.
  The chaos assumption holds near the point $h_x = h_z = 1$ at which I work in the rest of the paper,
  but breaks down in the free fermion limit $h_z = 0$, the paramagnetic limit $h_z \gg 1$,
  and in the presence of disorder.
  
  To check the chaos assumption I compute two quantities:
  a Hilbert space fraction for the eigenstates of the pool Hamiltonian,
  and the familiar gap ratio for their eigenenergies.
  The Hilbert space fraction directly measures how extended the eigenstates are.
  To define the Hilbert space fraction, start with the inverse participation ratio of the pool eigenstates $\ket{\alpha}$
  \begin{align}
    \mathcal I = \sum_{\bm \mu} |\braket{\alpha|\bm \mu}|^2\;.
  \end{align}
  For a Haar state this is $ \mathcal I_{\Haar} \approx 3/|\pool_l| $
  (in the limit of large Hilbert space dimension)
  because the fourth moment of the standard Gaussian is $3$.
  It is therefore useful to frame the IPR as the \textbf{Hilbert space fraction}
  \begin{align}
    \label{eq:H-frac}
    \mathcal F = |\pool_l| / \mathcal I\;:
  \end{align}
  this quantity has a straightforward interpretation as the fraction of the Hilbert space filled by the state.
  A Haar state has Hilbert space fraction
  \begin{align}
    \mathcal F_\Haar \approx \frac 1 3\;,
  \end{align}
  again in the limit of large Hilbert space dimension.
  The Hilbert space fraction \eqref{eq:H-frac} is a direct check of the chaos assumption Eq.~\eqref{eq:chaos}.

  The gap ratio indirectly diagnoses extended eigenstates by measuring level repulsion.
  To compute, sort the eigenenergies $E_j$ of the pool Hamiltonian, define gaps $\delta_j = E_{j+1} - E_j$, compute
  \begin{align}
   r_j = \frac{\min(\delta_j, \delta_{j+1})}{\max(\delta_j,\delta_{j+1})}\;,
  \end{align}
  and average over the spectrum.

  \subsubsection{Clean model}
  
  The Gaussian shape of the density of states is easily checked.
  Fig.~\ref{fig:gaussian-dos} shows the pool Hamiltonian density of states for the model \eqref{eq:tfim} at pool diameter $l = 7$.
  It follows a Gaussian past two standard deviations.
  
  The profusion of curves in Fig.~\ref{fig:gaussian-dos} obscures a curious notch and spike near $E = 0$.
  This is due to a small degenerate subspace exactly at $E = 0$ (within numerical precision),
  which may be related to powers of energy density.

  Fig.~\ref{fig:participation-fraction} top shows the participation fraction $\mathcal F$ as a function of longitudinal field $h_z$.
  For $h_z \lesssim 0.25 \lesssim 1.25$ the participation fraction is slightly less than the Haar value $\mathcal F_\Haar = 1/3$,
  but Fig.~\ref{fig:participation-fraction} bottom shows that it is does not decrease exponentially with pool size:
  the pool Hamiltonian eigenstates are extended.
  
  Fig.~\ref{fig:clean-gap-ratio} shows the gap ratio for $l = 7$ as a function of the longitudinal field $h_z$.
  (I have removed the small degenerate $E = 0$ subspace.)
  The gap ratio is close to its GOE value for  $0.25 \lesssim h_z \lesssim 1.25$,
  further indicating that the pool Hamiltonian is chaotic for those $h_z$.

  At the free fermion integrable point $h_z = 0$,
  the average Hilbert space fraction is small and the gap ratio is Poisson.
  This is because the dynamics conserves the number of $\sigma^{\pm} = \sigma^z \pm \sigma^y$ in a string---that is, the number of fermion operators.
  The pool Hilbert space therefore breaks up into sectors labeled by the number of fermion operators;
  even the largest sector is smaller than the whole pool by a combinatorial factor.

  At small but nonzero $h_z$, the $\sigma^z$ term gives a perturbative coupling between sectors,
  which hybridizes eigenstates in the different sectors.
  The participation fraction measures the resulting degree of hybridization.
  In the usual way the degree of hybridization is controlled by a competition between the coupling $h_z$ and the level spacing,
  which decreases exponentially with pool diameter $l$.
  This competition explains the non-monotonic behavior of the participation fraction $\mathcal F$ for $h_z \ll 1$.
  For small $l$, $\mathcal F$ decreases with $l$,
  because the dimension of the largest sector grows more slowly than the dimension of the whole pool;
  then it increases, as the sectors hybridize.

  In the limit of large $h_z$ the pool Hamiltonian becomes a paramagnet
  \begin{align}
    \label{eq:para-pool-ham}
    H_{\graph;l} \approx h_z \sum_j \Big(i\ketbra{\sigma^y_j}{\sigma^x_j} - i \ketbra{\sigma^x_j}{\sigma^y_j}\Big) \qquad (h_z \gg 1)
  \end{align}
  The eigenstates of this Hamiltonian are tensor product operators.
  Most eigenstates are $\sigma^z$ or $\sigma^0 = I$ on at least a few sites,
  so they have a low participation fraction.
  The pool Hamiltonian \eqref{eq:para-pool-ham} has many degeneracies (most obviously one can interchange $\sigma^z \leftrightarrow \sigma^0$ anywhere in the string, except at the ends, without changing the eigenvalue).
  This leads to a gap ratio below the Poisson value.
  At finite $h_z$ the transverse field and bond terms hybridize the eigenstates of this paramagnetic Hamiltonian,
  leading to larger participation ratio and a gap ratio that is nonzero, but may still be below the Poisson value.

  \begin{figure}
    \begin{minipage}{0.45\textwidth}
      \includegraphics[width=\textwidth]{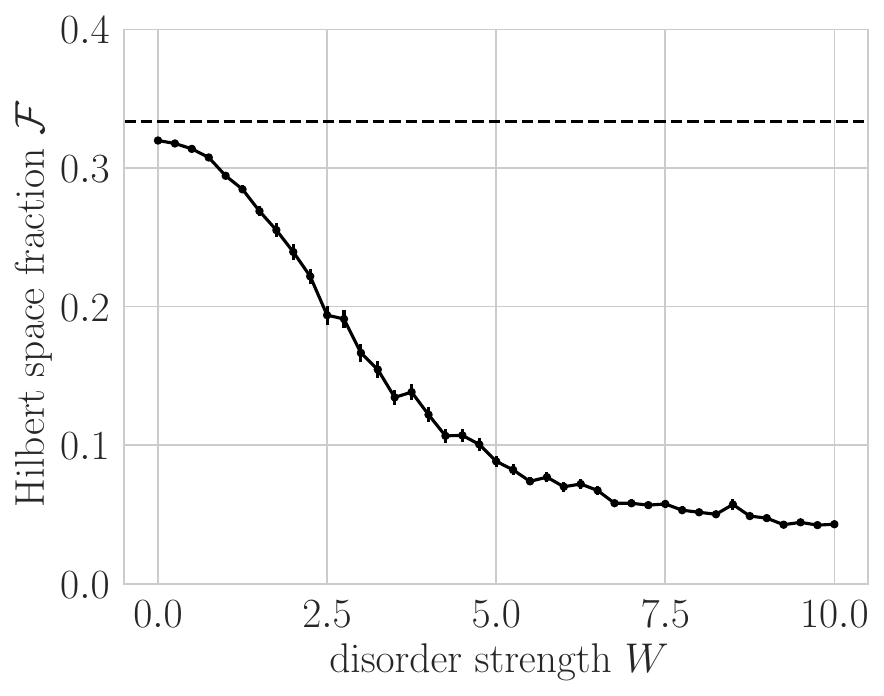}
    \end{minipage}
    \begin{minipage}{0.45\textwidth}
      \includegraphics[width=\textwidth]{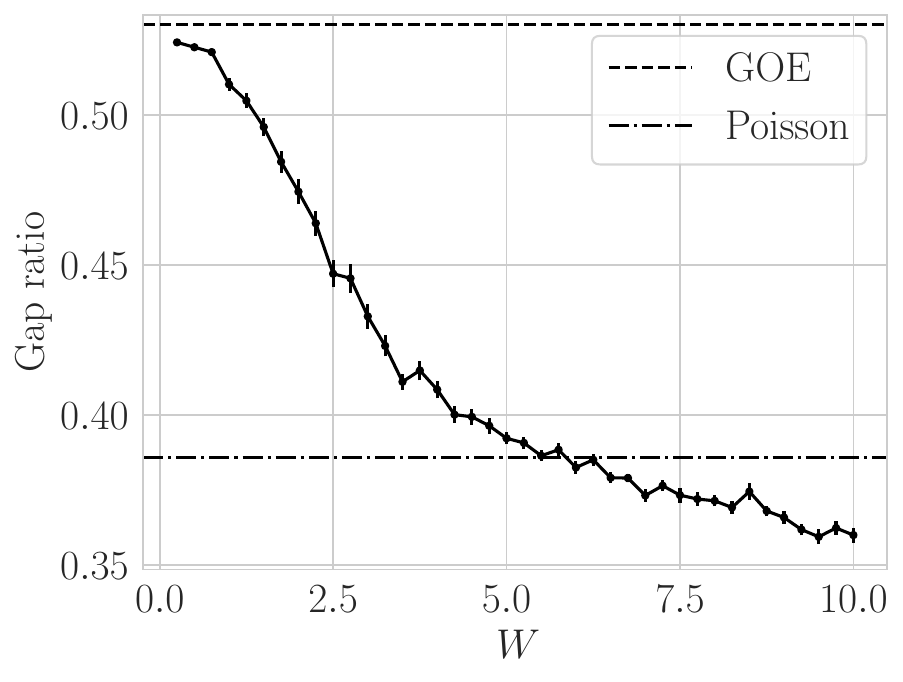}
    \end{minipage}
    \caption{
      \textbf{Disordered model:} eigenstate participation fraction (top) and gap ratio (bottom) for the pool Hamiltonian of the disordered model \eqref{eq:tfim-disorder} at pool diameter $l = 6$. I average over 100 disorder realizations. For small disorder the Hamiltonian pool is chaotic; for large disorder it is not.
      In the paramagnetic limit the gap ratio approaches $r = 0$ due degeneracies in the graph Hamiltonian for (cf Eq.~\eqref{eq:para-pool-ham} and discussion).
    }
    \label{fig:chaos-disordered}
  \end{figure}

  \subsubsection{Disordered model}

  Now add disorder to the model \eqref{eq:tfim} for
  \begin{align}
    \label{eq:tfim-disorder}
    H = \sum_j \sigma^z_j \sigma^z_{j+1} + \sum_j h_x \sigma^x_j + \sum_j (1 + \delta h^z_j) \sigma^z_j
  \end{align}
  where the $\delta h^z_j$ are drawn uniformly from $[-W,W]$.
  I choose the form $(1+ \delta h^z_j)$ for the field to avoid free-fermion integrability even at $W = 0$.

  Fig.~\ref{fig:chaos-disordered} shows the Hilbert space fraction (top) and gap ratio (bottom).
  Again, I have removed the small degenerate $E = 0$ subspace in calculating the gap ratio.
  The pool Hamiltonian is chaotic for small $W$, but trends towards a small Hilbert space fraction $\mathcal F$ and the paramagnetic gap ratio $r = 0$ as $W$ increases.
  
\section{Effective model for dynamics of small-diameter operators}\label{s:eff-model}
\begin{figure}
  \includegraphics[width=0.45\textwidth]{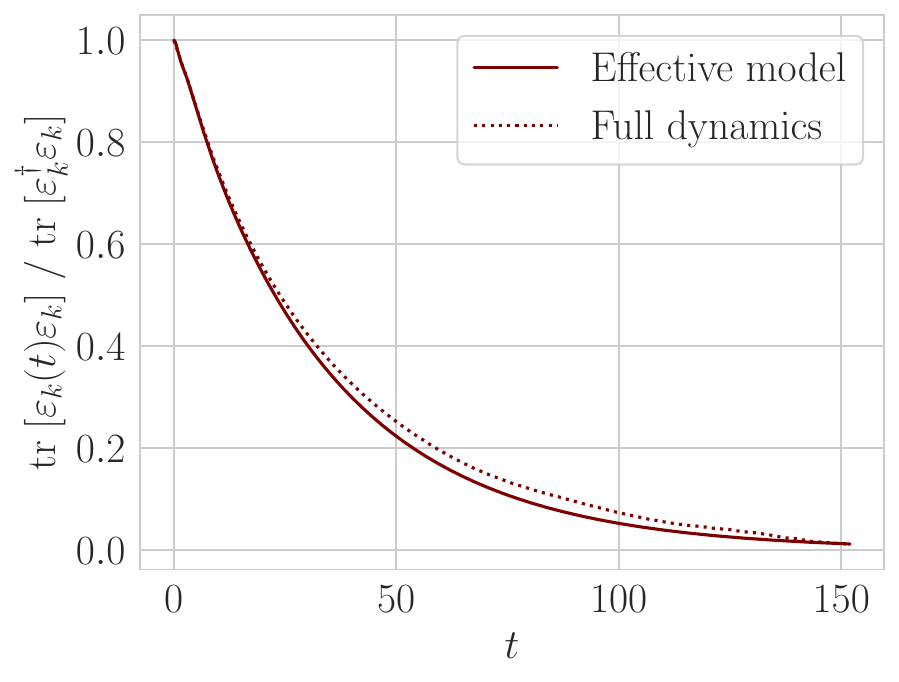}
  \caption{\textbf{Decay of Fourier cosine mode} Eq.~\eqref{eq:cosine-mode} in the effective model of Eq.~\eqref{eq:eff-model} (solid lines) and the full dynamics of the transverse field Ising model (dotted line) at system size $L = 20$. I use $l_* = 6$ and decay rate $\gamma_\semiemp = 4.49$.
    }
  \label{fig:fourier-decay}
\end{figure}

In Sec.~\ref{s:rate-loss} I considered hopping from $\bm \mu \in \pool_{l-1}$ to $\bm \lambda \in \pool_l$.
I argued that the rate of this hopping is controlled by the rate of further hopping to other sites in $\pool_l$.

Let us assume that for $l$ longer than some $l_*$, the particle does not return from such further hopping.
This is a nontrivial, uncontrolled approximation.
It essentially rules out the the possibility of ``backflow'', i.e. return of weight from long operators to short operators.
In App.~\ref{ss:fact-loss} I give some arguments and numerical evidence that this approximation is reasonable;
\onlinecite{vonkeyserlingkOperatorBackflowClassical2021} gives a more detailed calculation.

This no-backflow approximation is equivalent to replacing the dynamics on Pauli strings of diameter $l \ge l_*$ by a decay;
for simplicity, characterize the decay by a single decay rate
(a frequency-independent imaginary self-energy).

To be more precise, start with the graph Hamiltonian $H_{\graph}$.
Cut it off at a length scale $l_*$ by 
removing all the hoppings between operators of diameter $l \ge l_*+1$,
and endow them with a decay rate $\gamma$.
Since the model is nearest-neighbor, operators of diameter $\ge l_* + 2$ are isolated points---they do not affect the dynamics of any other operators.
The result of this procedure is an effective graph Hamiltonian
\begin{align}
  \label{eq:eff-model}
  \begin{split}
    H_{\graph;\mathrm{eff}}(\gamma,l_*) &= \sum_{\bm \mu, \diam \bm \nu \le l_*} [L_{\bm\mu\bm\nu}c^\dagger_{\bm\mu} c_{\bm\nu} + h.c.]\\
    &- i\gamma \sum_{\diam \bm \nu = l_* +1} c^\dagger_{\bm \nu} c_{\bm \nu}.\;.
  \end{split}
\end{align}
(Despite the second-quantized notation this model, like \eqref{eq:Hgraph}, is a single-particle model.)
This modified dynamics conserves the total energy.
In contrast to the exact dynamics, it does not conserve any of the powers of the energy:
$H^2$, for example,
has terms like $\varepsilon_1 \varepsilon_{L/2}$ that are long-ranged and subject to decay.

\begin{figure}[t]
  \includegraphics[width=.45\textwidth]{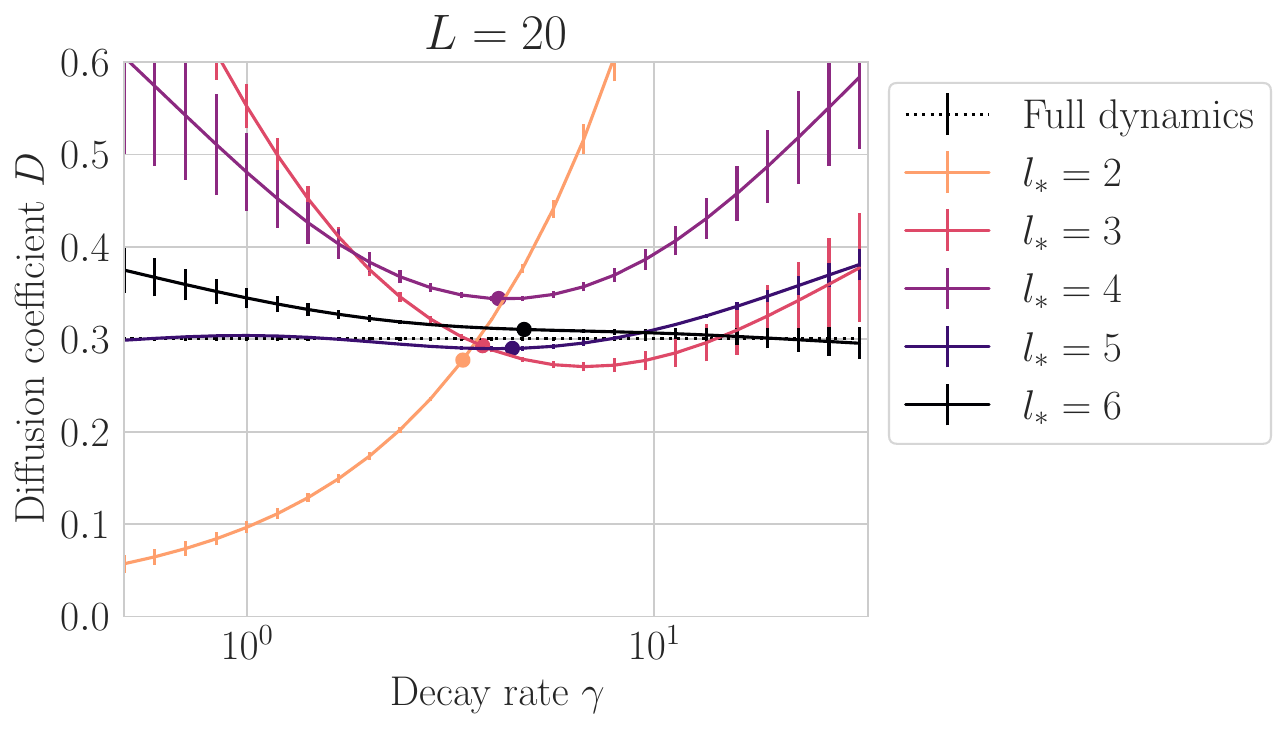}
  \caption{
    \textbf{ Diffusion coefficients in the effective model}
    as a function of $l_*$ (colors)
    and decay rate $\gamma$
     at system size $L = 20$.
    Colored balls show the semi-empirical decay rates $\gamma_{\semiemp}$;
    The dotted line shows a stochastic trace estimate of the exact dynamics.
    Error bars indicate the degree to which the dynamics are non-diffusive,
    complicating estimation of the diffusion coefficient (see App.~\ref{app:extract}).
    For the predicted decay rates,
    the effective model of Eq.~\eqref{eq:eff-model}
    shows cleanly diffusive behavior with diffusion coefficient very close to that of the exact dynamics.
  }
  \label{fig:diffusion}
\end{figure}

 When the Hamiltonian is translation invariant (as \eqref{eq:tfim} is),
 the effective model \eqref{eq:eff-model} commutes with translations,
 so it is block-diagonal in a basis of Fourier modes of Pauli strings.
 This leads to considerable savings  when simulating finite translation-invariant systems.
 It also makes possible simulation of infinite systems at no additional computational cost,
 but in this paper I restrict myself to finite systems for benchmarking purposes.

I simulate the decay of the longest Fourier cosine mode of the energy density
\begin{align}
  \label{eq:cosine-mode}
  \varepsilon_{k} = \sum_{j = 1}^L \cos(kj)
\end{align}
with $k = \frac{2\pi}{L}$ in the effective model \eqref{eq:eff-model} and the full dynamics \eqref{eq:tfim}.
(I use a stochastic trace to simulate the full dynamics;
see App.~\ref{app:full} for a brief description.)
Because the system is translation-invariant,
\begin{align}
  \expct{\varepsilon_{k'}\varepsilon_k(t)} = \tr[\varepsilon_{k'}\varepsilon_k(t)] \propto \delta_{kk'}\;.
\end{align}
I plot $\expct{\varepsilon_k\varepsilon_k(t)}$ for $L = 20, l_* = 6, \gamma = 4 \approx \gamma_{\semiemp}$ in Fig.~\eqref{fig:fourier-decay};
the effective model agrees well with the full dynamics.
(See App.~\ref{app:full} for a description of the simulation of the full dynamics.)

\begin{figure}[t]
  \includegraphics[width=.45\textwidth]{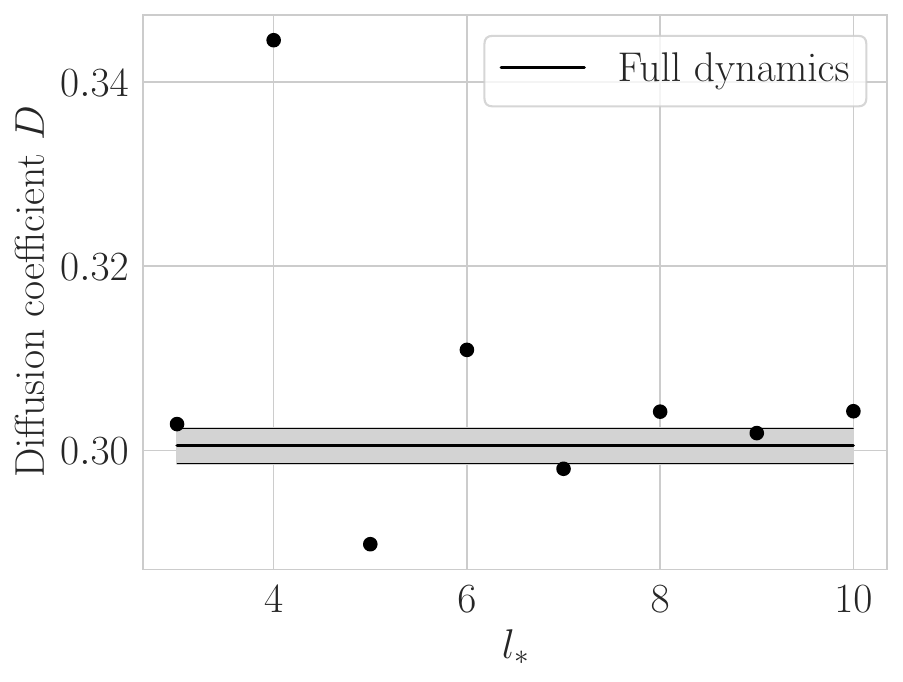}
  
  \includegraphics[width=.45\textwidth]{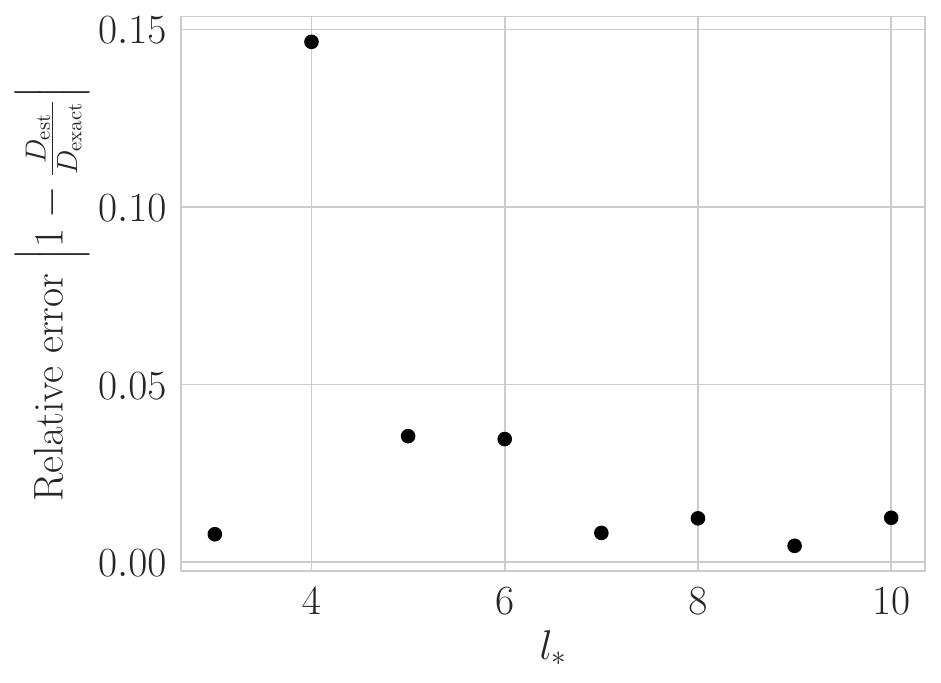}
  \caption{
    \textbf{ Diffusion coefficients in the effective model at predicted decay rate $\gamma_{\semiemp}$ } (black balls) as a function of cutoff length $l_*$,
    together with the diffusion coefficient estimated from the full dynamics
    (black line; shaded region indicates uncertainty).
    The largest-$l_*$ value differs from the full dynamics by $\approx 1.25 \%$.
  }
  \label{fig:diffusion-lstar}
\end{figure}

\begin{figure*}[t]
  \centering\includegraphics[width=\textwidth]{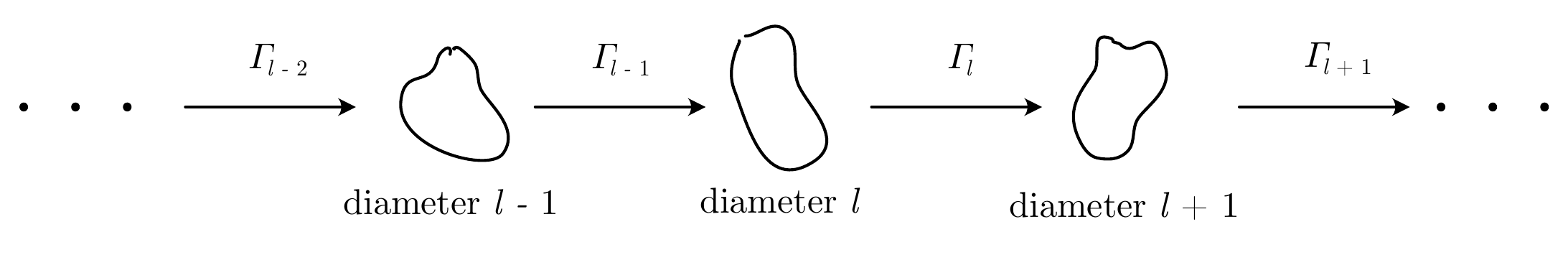}
  \caption{
    \textbf{Caricature of Liouvillian graph dynamics:} the pool of operators of diameter $l$ is connected to the pool of diameter $l+1$ by an escape rate $\Gamma_l$.
  }
  \label{fig:pool-chain-caricature}
\end{figure*}

Fig.~\ref{fig:diffusion} shows diffusion coefficients extracted from the decay of the Fourier mode \eqref{eq:cosine-mode}
as a function of $l_*$ and $\gamma$;
the error bars give a heuristic indication of how non-diffusive the dynamics are.
(App.~\ref{app:extract} gives details of the extraction of diffusion coefficients
and of this measure of how non-diffusive the dynamics are.)
For $\gamma = \gamma_{\semiemp}$ (marked by the colored balls),
the effective dynamics gives cleanly diffusive dynamics and a diffusion coefficient that agrees well with the full dynamics,
with the exception of $l_* = 5$.

Fig.~\ref{fig:diffusion-lstar} I show the diffusion coefficient the semi-empirical decay rate $\gamma_{\semiemp}$ as a function of $l_*$.
The estimated diffusion coefficient alternates before appearing to converge to a value $\approx 1 \%$ greater than the diffusion coefficient in the full dynamics.

The large-$l_*$ discrepancy may be explained by the imprecision in my estimate of the diffusion coefficient for the full dynamics,
but this is unlikely:
the $l_* = 10$ value lies outside the error bars on the estimate from the full dynamics,
and it appears that the value about which the estimated diffusion coefficients alternate likewise lies outside those error bars.
It is more likely that the discrepancy comes from finite size effects in the full dynamics (performed on a system of size $L = 20$), or from coherent effects not included in the effective model \eqref{eq:eff-model}.

The alternation comes about because the semi-empirical decay rate $\gamma_{\semiemp}$ does not match the true decay rates.
To understand this alternation,
characterize the dynamics as a chain of pools of fixed operator diameter, connected by escape rates $\Gamma_l$ (Fig.~\ref{fig:pool-chain-caricature});
these escape rates measure how fast an operator in pool $\pool_{l-1}$ escapes to pool $\pool_{l}$.
As in the toy model of Sec.~\ref{ss:loss-toy},
the escape rate is controlled by a matrix element $V_{l-1}$ between sites in $\pool_{l-1}$ and sites in $\pool_{l}$,
together with the decay rate $\gamma_{l}$ in $\pool_{l}$:
\begin{align}
  \Gamma_{l-1} \sim |V_{l-1}|^2 / \gamma_{l}\;.
\end{align}
(The dependence on $\gamma_{l}$ comes about due to a quantum Zeno effect.)
But the decay rate $\gamma_{l}$ has two components:
hopping to $O(l)$ other sites in $\pool_{l}$,
and hopping to $O(1)$ sites in $\pool_{l+1}$.
The hopping to $\pool_{l+1}$ motivates the semi-empirical correction $\sqrt{l(J^2 + h_x^2 + h_z^2)} \to\sqrt{(l+1)(J^2 + h_x^2 + h_z^2)}$,
but the replacement $l \to l+1$ crudely simplifies the effect.
More precisely, one expects something along the lines of
\begin{align}
  \gamma_l = \gamma_{l;\text{intrinsic} } + \Gamma_{l}\;.
\end{align}
In this picture the effective model \eqref{eq:eff-model} corresponds to cutting off the chain after pool $\pool_{l_*+1}$
and replacing $\gamma_{l_*+1} \to \gamma'_{l_*+1}$,
where we control $\gamma'$.

Now suppose we pick $\gamma'_{l_*+1}$ incorrectly:
\begin{align}
  \gamma' = \gamma_{l_*+1} + \Delta \gamma\;.
\end{align}
Then the effective model has
\begin{align}
  \Gamma'_{l_*} \sim  |V_{l_*}|^2 / \gamma'_{l_*+1} \approx \Gamma_{l_*}\left(1 - \frac{\Delta\gamma}{\gamma_{l_*+1}}\right)
\end{align}
and
\begin{align}
  \gamma_{l_*}' &= \gamma_{l_*} - \frac{\Gamma_{l_*}}{\gamma_{l_*+1}} \Delta \gamma\;.
\end{align}
So an error $+\Delta \gamma$ in the decay rate for the pool of diameter $l_*+1$
corresponds to an error $\propto -\Delta \gamma$ in the decay rate for the pool of diameter $l_*$.
Since $\frac d {d\gamma}  D_{\mathrm{est}} \ne 0$, this alternation in the decay rate gives an alternation in $D_{\mathrm est}$.

  These numerics are not sensitive enough to detect long-time tails---or failures of the effective model \eqref{eq:eff-model} to capture those long-time tails.
  Such behavior will be hidden in imperfections in the scaling collapses I discuss in App.~\ref{app:extract}.
}
\section{Discussion}\label{s:discussion}

I have argued that the structure of the Liouvillian graph determines the behavior of thermalizing Hamiltonians
and gives a non-Hermitian effective model for the dynamics of small-diameter operators.
I first followed \onlinecite{altman_ehud_computing_2018,parker_universal_2019,cao_statistical_2021} in mapping Heisenberg dynamics on operators
to single-particle hopping on a graph whose vertices are Pauli matrices.
I noted that the Liouvillian graph divides naturally into ``pools'' of fixed left and right endpoints and consequently fixed diameter;
these pools are thickly intraconnected but thinly interconnected.
I then followed \onlinecite{khemani_operator_2018} in understanding the effective non-unitarity of diffusion 
as escape of particles into a ``vacuum'' of long operators,
and argued that the escape rate is controlled by the coherent dynamics of the fixed-diameter pools.
This argument, which assumes that dynamics in a pool is chaotic,
gives a quantitative prediction for the escape rate,
which I verified numerically.
The Liouvillian graph picture also leads to a non-Hermitian effective model
in which dynamics on operators longer than some $l_*$ is replaced by a decay, with decay rate predicted by the coherent pool dynamics.
This effective model gives good agreement with the full Hamiltonian dynamics for the Ising model \eqref{eq:tfim}.

  The effective model of Sec.~\ref{s:eff-model} is, in essence, a poor man's memory kernel model \cite{forster_hydrodynamic_1975,zwanzig_memory_1961,mori_transport_1965},
  in which I have projected out operators of diameter $\diam \bm \nu \ge l_* + 1$.
  From this perspective Sec.~\ref{ss:rate-loss-est} is an argument that
  the memory kernel is a Gaussian with rate $\gamma$;
  Sec.~\ref{s:eff-model} replaces that Gaussian by an exponential decay with rate $\gamma_{\semiemp}$.
  This turns the integro-differential equation given by the Gaussian memory kernel into a system of linear first-order ODEs
  (specified by Eq.~\eqref{eq:eff-model})
  on a somewhat larger Hilbert space.
  Other projection operator techniques \cite{grabert_projection_1982} may offer better effective models.

  The effective model is less closely related to recursion approaches
  (see \onlinecite{viswanath_recursion_1994} for a review).
  There one reconstructs the quantities of interest from moments
  \begin{equation}
    \label{eq:moment}
    M_n = \tr\Big(A [H,[H,\dots, [H,A]\dots]]\Big)
  \end{equation}
  of some quantity of interest $A$ or, equivalently, coefficients in a continued-fraction expansion for a Laplace transform of the dynamics of $A$.
  In either case one can only compute finitely many coefficients.
  Straightforward truncation corresponds to a short-time expansion,
  which may converge to the correct long-time limit.
  \footnote{
      I am a priori skeptical that this convergence is trustworthy.
      The authors of \onlinecite{viswanath_recursion_1994} are likewise skeptical,
      and in Sec.~8.1 they argue that the pole structure of a response function depends on all the continued-fraction coefficients (hence all the moments $M_n$).
      But the numerics of \onlinecite{viswanath_recursion_1994} Sec.~8.1 are marred by the fact they work at finite system size,
      rather than working directly in Fourier space as in \onlinecite{jung_transport_2006}.
      \onlinecite{jung_transport_2006} performs simple truncation at order $n \sim 20$,
      but then extrapolates the resulting DC conductivity to $n = \infty$.
      This is no worse than most other numerical methods, which likewise rely on convergence testing,
      and the result of \onlinecite{jung_transport_2006}---that certain $SU(2)$-symmetric perturbations to the Heisenberg model do \textit{not} break anomalous diffusion---is consistent with very different recent work \cite{de_nardis_stability_2021}.
      I am curious (1) whether the extrapolation can be physically justified, and (2) whether it gives correct diffusion coefficients in non-integrable systems
  }
  But one would desire a thoughtful strategy for extracting the relevant information from low moments,
  which may come in the form of a physically-motivated ansatz for the behavior of all moments.
  The universal operator growth hypothesis of \onlinecite{parker_universal_2019} is just such an ansatz.

  The effective model of Sec.~\ref{s:eff-model} bears roughly the same relationship to simple truncation of a moment expansion
  that a cluster expansion for a Gibbs state $e^{-\beta H}$
  (as in \onlinecite{vanhecke_simulating_2021})
  bears to a simple a simple Taylor series expansion.
  Computing the dynamics of the effective model \eqref{eq:eff-model} by exact diagonalization is analogous to working with \textit{arbitrarily high} moments of the form \eqref{eq:moment}%
  ---but these moments are modified to only take into account terms containing operators up to diameter $l_*$.
  So, while it may be possible to rephrase the effective model of Sec.~\ref{s:eff-model} as a truncation strategy for a recursion method,
  this will not be straightforward.

The effective model is useful in its own right.
Its Hilbert space dimension is
\begin{equation}
  N_{\mathrm{eff}} \sim L 4^{l_*}\;,
\end{equation}
so memory and computation time requirements
scale exponentially in the length scale $l_*$,
 which is broadly speaking an IR cutoff for the effective model and a UV cutoff for diffusion.
but polynomially, indeed linearly, in system size.
The resource requirements for simulations are therefore controlled primarily by $l_*$;
for models in which small $l_*$ are suitable---%
i.e. models in which even small pools are chaotic in the sense required by section \ref{ss:rate-loss-est}---%
the required Hilbert space will be small.

The effective model also offers the exciting prospect of two- and higher-dimensional simulations.
The construction of the effective model rests on the decomposition of the Liouvillian graph into pools of fixed diameter,
which in turn rests on the decreasing surface-area-to-volume ratio of these pools.
Higher dimensions will show the same graph decomposition.
Roughly, in dimension $d$
an operator of diameter $l$
only connects to operators of diameter $l-1$ or $l+1$
via the $O(l^{d-1})$ Hamiltonian terms at its edges,
but connects to other operators of diameter $l$ via the $O(l^d)$ Hamiltonian terms in its interior.
(Defining diameter in this case will not be entirely straightforward:
one will want to take into account the underlying lattice geometry in defining the ``pools'').

I close by emphasizing some limitations of the effective model and the Liouvillian graph picture more generally.

First, the effective model is not a satisfactory replacement for existing numerical methods
(DMT \cite{white_quantum_2018,ye_emergent_2019}, DAOE \cite{rakovszky_dissipation-assisted_2020}, and ordinary TEBD in non-equilibrium steady state experiments).
The effective model has only one free parameter,
the length scale $l_*$.
For each $l_*$ there is a unique correct decay rate;
the bulk of this paper has been devoted to estimating that rate.
Convergence testing therefore
requires increasing $l_*$, which is exponentially time-consuming.
DMT and DAOE, by contrast, are parameterized not only by a length scale like $l_*$
but also by another continuously tunable accuracy parameter:
the bond dimension in DMT and the decay rate for DAOE.
Both methods can thus accurately treat at least some long-range correlations
on at least some timescales---and, importantly, one can use these parameters to smoothly approach the exact limit.

Second, the effective model assumes that the pool Hamiltonians are chaotic---but many spin systems do not have this property.
Free-fermion systems, Bethe ansatz integrable systems, and many-body localized systems
never thermalize,
and weak integrability breaking leads to thermalization only on long timescales.
In each case the failure of thermalization corresponds to a breakdown of either the graph structure of Sec.~\ref{ss:graph-structure} or the chaos assumption of Sec.~\ref{ss:rate-loss-est}.
 Sec.~\ref{ss:chaos-breakdown} discusses this question for localized and free fermion integrable models;
for Bethe ansatz integrable models
the tower of local conserved quantities will strongly constrain the pool dynamics.

Third, these Liouvillian graph calculations require a high-temperature starting state,
for the same reason that DAOE does.
The expectation value of an operator $A(t)$ in an initial state $\rho$ is
\[ \expct{A} = \tr A \rho\;. \]
The effective model lets us estimate the projection of the Heisenberg evolution $A(t)$
of some operator onto short Pauli strings.
But if the initial state $\rho$ is not close to the infinite-temperature density matrix,
it can have appreciable weight on Pauli strings of long diameter.
This is true even if the initial state has some short correlation length $\xi$:
if $|i - j| \gg \xi$,
\[ \expct{\sigma^z_i \sigma^z_j} \approx \expct{\sigma^z_i}\expct{\sigma^z_j}\;. \]
One could treat these correlations by means of a nonlinear effective evolution of short-diameter correlations,
in which longer correlations are replaced by these ``classical correlations'',
as in DMT.
The method of \onlinecite{kvorning_time-evolution_2021} is a major step in this direction,
and in fact goes far beyond the present work in its potential for treating finite-temperature dynamics.

% \subsection{Fluctuations}
% In the introduction I situated DAOE, DMT, etc. as a \textit{Stosszahlansatz}, a ``molecular chaos assumption'' that breaks unitarity by discarding correlations.
% But we can understand from alternate point of view.
% ``Long operators'' measure not only correlations but fluctuations.
% (Heisenberg formulation / density matrix treats statistical and quantum fluctuations on same footing?)

% \onlinecite{andreev_corrections_1978}. Maybe \onlinecite{hohenberg_theory_1977}? Stephanov's hydro+ work isn't so relevant, but his PRC 100, 024910 ``Relativistic Hydrodynamic Fluctuations'' maybe is.

\begin{acknowledgements}
  %This work is heavily indebted to work of Ehud Altman, Xiangyu Cao, Arnold Mong, and David Huse presented at the summer 2018 KITP conference ``Novel Approaches to Quantum Dynamics'', and to conversations with Altman about that work.

  I am grateful to Brian Swingle, Yong-Chan Yoo, Michael Buchhold, Kai Klocke, and Kevin Slagle for helpful conversations and commentary on this work, and to Brayden Ware, Stuart Thomas, and Jay Deep Sau for helpful conversations in the context of related work.

  I gratefully acknowledge the U.S. Department of Energy (DOE), Office of Science, Office of Advanced Scientific Computing Research (ASCR) Quantum Computing Application Teams program, for support under fieldwork proposal number ERKJ347.
\end{acknowledgements}

\bibliography{references}

\begin{thebibliography}{58}%
\makeatletter
\providecommand \@ifxundefined [1]{%
 \@ifx{#1\undefined}
}%
\providecommand \@ifnum [1]{%
 \ifnum #1\expandafter \@firstoftwo
 \else \expandafter \@secondoftwo
 \fi
}%
\providecommand \@ifx [1]{%
 \ifx #1\expandafter \@firstoftwo
 \else \expandafter \@secondoftwo
 \fi
}%
\providecommand \natexlab [1]{#1}%
\providecommand \enquote  [1]{``#1''}%
\providecommand \bibnamefont  [1]{#1}%
\providecommand \bibfnamefont [1]{#1}%
\providecommand \citenamefont [1]{#1}%
\providecommand \href@noop [0]{\@secondoftwo}%
\providecommand \href [0]{\begingroup \@sanitize@url \@href}%
\providecommand \@href[1]{\@@startlink{#1}\@@href}%
\providecommand \@@href[1]{\endgroup#1\@@endlink}%
\providecommand \@sanitize@url [0]{\catcode `\\12\catcode `\$12\catcode
  `\&12\catcode `\#12\catcode `\^12\catcode `\_12\catcode `\%12\relax}%
\providecommand \@@startlink[1]{}%
\providecommand \@@endlink[0]{}%
\providecommand \url  [0]{\begingroup\@sanitize@url \@url }%
\providecommand \@url [1]{\endgroup\@href {#1}{\urlprefix }}%
\providecommand \urlprefix  [0]{URL }%
\providecommand \Eprint [0]{\href }%
\providecommand \doibase [0]{http://dx.doi.org/}%
\providecommand \selectlanguage [0]{\@gobble}%
\providecommand \bibinfo  [0]{\@secondoftwo}%
\providecommand \bibfield  [0]{\@secondoftwo}%
\providecommand \translation [1]{[#1]}%
\providecommand \BibitemOpen [0]{}%
\providecommand \bibitemStop [0]{}%
\providecommand \bibitemNoStop [0]{.\EOS\space}%
\providecommand \EOS [0]{\spacefactor3000\relax}%
\providecommand \BibitemShut  [1]{\csname bibitem#1\endcsname}%
\let\auto@bib@innerbib\@empty
%</preamble>
\bibitem [{\citenamefont {Forster}(1975)}]{forster_hydrodynamic_1975}%
  \BibitemOpen
  \bibfield  {author} {\bibinfo {author} {\bibfnamefont {D.}~\bibnamefont
  {Forster}},\ }\href@noop {} {\emph {\bibinfo {title} {{Hydrodynamic
  Fluctuations, Broken Symmetry, and Correlation Functions}}}},\ Frontiers in
  physics ; 47\ (\bibinfo  {publisher} {W.A. Benjamin, Advanced Book Program},\
  \bibinfo {address} {Reading, Mass.},\ \bibinfo {year} {1975})\ \bibinfo
  {note} {section: xix, 326 pages : illustrations ; 25 cm.}\BibitemShut {Stop}%
\bibitem [{\citenamefont {Chaikin}\ and\ \citenamefont
  {Lubensky}(2000)}]{chaikin_principles_2000}%
  \BibitemOpen
  \bibfield  {author} {\bibinfo {author} {\bibfnamefont {P.~M.}\ \bibnamefont
  {Chaikin}}\ and\ \bibinfo {author} {\bibfnamefont {T.~C.}\ \bibnamefont
  {Lubensky}},\ }\href@noop {} {\emph {\bibinfo {title} {Principles of
  {Condensed} {Matter} {Physics}}}}\ (\bibinfo  {publisher} {Cambridge
  University Press},\ \bibinfo {address} {Cambridge},\ \bibinfo {year}
  {2000})\BibitemShut {NoStop}%
\bibitem [{Note1()}]{Note1}%
  \BibitemOpen
  \bibinfo {note} {Higher-order corrections to the diffusion equation \protect
  \textup {\hbox {\mathsurround \z@ \protect \normalfont (\ignorespaces \ref
  {eq:diff}\unskip \@@italiccorr )}} can give nontrivial physics, including
  ``long-time tails''\cite
  {kovtun_hydrodynamic_2003,mukerjee_towards_2006,spohn_nonlinear_2014,lux_hydrodynamic_2014,kirkpatrick_non-hydrodynamic_2021}.
  This work treats the microscopic physics leading to the diffusion equation
  (with or without corrections) rather than the physics described by that
  equation---but the interplay between that microscopic physics and the
  long-time tails, analogous to the interplay investigated for the Boltzmann
  equation in \protect \rev@citealpnum {kirkpatrick_non-hydrodynamic_2021}, may
  be an interesting future line of inquiry.}\BibitemShut {Stop}%
\bibitem [{Note2()}]{Note2}%
  \BibitemOpen
  \bibinfo {note} {Our understanding of the many body localization transition
  rests on phenomenological renormalization group theories that take for
  granted the emergence of hydrodynamics from sufficiently strongly-interacting
  systems.}\BibitemShut {Stop}%
\bibitem [{\citenamefont {Potter}\ \emph {et~al.}(2015)\citenamefont {Potter},
  \citenamefont {Vasseur},\ and\ \citenamefont
  {Parameswaran}}]{potter_universal_2015}%
  \BibitemOpen
  \bibfield  {author} {\bibinfo {author} {\bibfnamefont {A.~C.}\ \bibnamefont
  {Potter}}, \bibinfo {author} {\bibfnamefont {R.}~\bibnamefont {Vasseur}}, \
  and\ \bibinfo {author} {\bibfnamefont {S.}~\bibnamefont {Parameswaran}},\
  }\href {\doibase 10.1103/PhysRevX.5.031033} {\bibfield  {journal} {\bibinfo
  {journal} {Physical Review X}\ }\textbf {\bibinfo {volume} {5}},\ \bibinfo
  {pages} {031033} (\bibinfo {year} {2015})}\BibitemShut {NoStop}%
\bibitem [{\citenamefont {Vosk}\ \emph {et~al.}(2015)\citenamefont {Vosk},
  \citenamefont {Huse},\ and\ \citenamefont {Altman}}]{vosk_theory_2015}%
  \BibitemOpen
  \bibfield  {author} {\bibinfo {author} {\bibfnamefont {R.}~\bibnamefont
  {Vosk}}, \bibinfo {author} {\bibfnamefont {D.~A.}\ \bibnamefont {Huse}}, \
  and\ \bibinfo {author} {\bibfnamefont {E.}~\bibnamefont {Altman}},\ }\href
  {\doibase 10.1103/PhysRevX.5.031032} {\bibfield  {journal} {\bibinfo
  {journal} {Physical Review X}\ }\textbf {\bibinfo {volume} {5}},\ \bibinfo
  {pages} {031032} (\bibinfo {year} {2015})}\BibitemShut {NoStop}%
\bibitem [{\citenamefont {Dumitrescu}\ \emph {et~al.}(2017)\citenamefont
  {Dumitrescu}, \citenamefont {Vasseur},\ and\ \citenamefont
  {Potter}}]{dumitrescu_scaling_2017}%
  \BibitemOpen
  \bibfield  {author} {\bibinfo {author} {\bibfnamefont {P.~T.}\ \bibnamefont
  {Dumitrescu}}, \bibinfo {author} {\bibfnamefont {R.}~\bibnamefont {Vasseur}},
  \ and\ \bibinfo {author} {\bibfnamefont {A.~C.}\ \bibnamefont {Potter}},\
  }\href {\doibase 10.1103/PhysRevLett.119.110604} {\bibfield  {journal}
  {\bibinfo  {journal} {Physical Review Letters}\ }\textbf {\bibinfo {volume}
  {119}},\ \bibinfo {pages} {110604} (\bibinfo {year} {2017})},\ \bibinfo
  {note} {arXiv: 1701.04827}\BibitemShut {NoStop}%
\bibitem [{\citenamefont {Goremykina}\ \emph {et~al.}(2019)\citenamefont
  {Goremykina}, \citenamefont {Vasseur},\ and\ \citenamefont
  {Serbyn}}]{goremykina_analytically_2018}%
  \BibitemOpen
  \bibfield  {author} {\bibinfo {author} {\bibfnamefont {A.}~\bibnamefont
  {Goremykina}}, \bibinfo {author} {\bibfnamefont {R.}~\bibnamefont {Vasseur}},
  \ and\ \bibinfo {author} {\bibfnamefont {M.}~\bibnamefont {Serbyn}},\ }\href
  {\doibase 10.1103/physrevlett.122.040601} {\bibfield  {journal} {\bibinfo
  {journal} {Physical Review Letters}\ }\textbf {\bibinfo {volume} {122}}
  (\bibinfo {year} {2019}),\ 10.1103/physrevlett.122.040601}\BibitemShut
  {NoStop}%
\bibitem [{\citenamefont {Morningstar}\ and\ \citenamefont
  {Huse}(2019)}]{morningstar_renormalization-group_2019}%
  \BibitemOpen
  \bibfield  {author} {\bibinfo {author} {\bibfnamefont {A.}~\bibnamefont
  {Morningstar}}\ and\ \bibinfo {author} {\bibfnamefont {D.~A.}\ \bibnamefont
  {Huse}},\ }\href {\doibase 10.1103/physrevb.99.224205} {\bibfield  {journal}
  {\bibinfo  {journal} {Physical Review B}\ }\textbf {\bibinfo {volume} {99}}
  (\bibinfo {year} {2019}),\ 10.1103/physrevb.99.224205}\BibitemShut {NoStop}%
\bibitem [{\citenamefont {Morningstar}\ \emph {et~al.}(2020)\citenamefont
  {Morningstar}, \citenamefont {Huse},\ and\ \citenamefont
  {Imbrie}}]{morningstar_many-body_2020}%
  \BibitemOpen
  \bibfield  {author} {\bibinfo {author} {\bibfnamefont {A.}~\bibnamefont
  {Morningstar}}, \bibinfo {author} {\bibfnamefont {D.~A.}\ \bibnamefont
  {Huse}}, \ and\ \bibinfo {author} {\bibfnamefont {J.~Z.}\ \bibnamefont
  {Imbrie}},\ }\href {\doibase 10.1103/physrevb.102.125134} {\bibfield
  {journal} {\bibinfo  {journal} {Physical Review B}\ }\textbf {\bibinfo
  {volume} {102}} (\bibinfo {year} {2020}),\
  10.1103/physrevb.102.125134}\BibitemShut {NoStop}%
\bibitem [{\citenamefont {De~Roeck}\ and\ \citenamefont
  {Huveneers}(2017)}]{de_roeck_stability_2017}%
  \BibitemOpen
  \bibfield  {author} {\bibinfo {author} {\bibfnamefont {W.}~\bibnamefont
  {De~Roeck}}\ and\ \bibinfo {author} {\bibfnamefont {F.}~\bibnamefont
  {Huveneers}},\ }\href {\doibase 10.1103/PhysRevB.95.155129} {\bibfield
  {journal} {\bibinfo  {journal} {Physical Review B}\ }\textbf {\bibinfo
  {volume} {95}} (\bibinfo {year} {2017}),\ 10.1103/PhysRevB.95.155129},\
  \bibinfo {note} {arXiv: 1608.01815}\BibitemShut {NoStop}%
\bibitem [{\citenamefont {Thiery}\ \emph {et~al.}(2018)\citenamefont {Thiery},
  \citenamefont {Huveneers}, \citenamefont {Müller},\ and\ \citenamefont
  {Roeck}}]{thiery_many-body_2017}%
  \BibitemOpen
  \bibfield  {author} {\bibinfo {author} {\bibfnamefont {T.}~\bibnamefont
  {Thiery}}, \bibinfo {author} {\bibfnamefont {F.}~\bibnamefont {Huveneers}},
  \bibinfo {author} {\bibfnamefont {M.}~\bibnamefont {Müller}}, \ and\
  \bibinfo {author} {\bibfnamefont {W.~D.}\ \bibnamefont {Roeck}},\ }\href
  {\doibase 10.1103/physrevlett.121.140601} {\bibfield  {journal} {\bibinfo
  {journal} {Physical Review Letters}\ }\textbf {\bibinfo {volume} {121}}
  (\bibinfo {year} {2018}),\ 10.1103/physrevlett.121.140601}\BibitemShut
  {NoStop}%
\bibitem [{\citenamefont {Thiery}\ \emph {et~al.}(2017)\citenamefont {Thiery},
  \citenamefont {Müller},\ and\ \citenamefont
  {De~Roeck}}]{thiery_microscopically_2017}%
  \BibitemOpen
  \bibfield  {author} {\bibinfo {author} {\bibfnamefont {T.}~\bibnamefont
  {Thiery}}, \bibinfo {author} {\bibfnamefont {M.}~\bibnamefont {Müller}}, \
  and\ \bibinfo {author} {\bibfnamefont {W.}~\bibnamefont {De~Roeck}},\ }\href
  {http://arxiv.org/abs/1711.09880} {\bibfield  {journal} {\bibinfo  {journal}
  {arXiv:1711.09880 [cond-mat]}\ } (\bibinfo {year} {2017})},\ \bibinfo {note}
  {arXiv: 1711.09880}\BibitemShut {NoStop}%
\bibitem [{Note3()}]{Note3}%
  \BibitemOpen
  \bibinfo {note} {Heavy-ion collisions produce high-temperature clouds of
  quarks and gluons. As these clouds expand and cool, they are pass through an
  expected line of first-order phase transitions that terminate in a ``critical
  end point''. Statistical fluctuation ``freeze out'' via the Kibble-Zurek
  mechanism when the cloud passes near the critical endpoint, leading to
  experimentally visible signatures of the details of the transition. But
  detailed comparison of experiments to theory requires first-principles
  calculations of hydrodynamical transport coefficients.}\BibitemShut {Stop}%
\bibitem [{\citenamefont {Stephanov}\ \emph {et~al.}(1998)\citenamefont
  {Stephanov}, \citenamefont {Rajagopal},\ and\ \citenamefont
  {Shuryak}}]{stephanov_signatures_1998}%
  \BibitemOpen
  \bibfield  {author} {\bibinfo {author} {\bibfnamefont {M.}~\bibnamefont
  {Stephanov}}, \bibinfo {author} {\bibfnamefont {K.}~\bibnamefont
  {Rajagopal}}, \ and\ \bibinfo {author} {\bibfnamefont {E.}~\bibnamefont
  {Shuryak}},\ }\href {\doibase 10.1103/PhysRevLett.81.4816} {\bibfield
  {journal} {\bibinfo  {journal} {Physical Review Letters}\ }\textbf {\bibinfo
  {volume} {81}},\ \bibinfo {pages} {4816} (\bibinfo {year} {1998})},\ \bibinfo
  {note} {arXiv: hep-ph/9806219}\BibitemShut {NoStop}%
\bibitem [{\citenamefont {Kolb}\ \emph {et~al.}(2000)\citenamefont {Kolb},
  \citenamefont {Sollfrank},\ and\ \citenamefont
  {Heinz}}]{kolb_anisotropic_2000}%
  \BibitemOpen
  \bibfield  {author} {\bibinfo {author} {\bibfnamefont {P.~F.}\ \bibnamefont
  {Kolb}}, \bibinfo {author} {\bibfnamefont {J.}~\bibnamefont {Sollfrank}}, \
  and\ \bibinfo {author} {\bibfnamefont {U.}~\bibnamefont {Heinz}},\ }\href
  {\doibase 10.1103/PhysRevC.62.054909} {\bibfield  {journal} {\bibinfo
  {journal} {Physical Review C}\ }\textbf {\bibinfo {volume} {62}},\ \bibinfo
  {pages} {054909} (\bibinfo {year} {2000})},\ \bibinfo {note} {arXiv:
  hep-ph/0006129}\BibitemShut {NoStop}%
\bibitem [{\citenamefont {Ollitrault}(1992)}]{ollitrault_anisotropy_1992}%
  \BibitemOpen
  \bibfield  {author} {\bibinfo {author} {\bibfnamefont {J.-Y.}\ \bibnamefont
  {Ollitrault}},\ }\href {\doibase 10.1103/PhysRevD.46.229} {\bibfield
  {journal} {\bibinfo  {journal} {Physical Review D}\ }\textbf {\bibinfo
  {volume} {46}},\ \bibinfo {pages} {229} (\bibinfo {year} {1992})}\BibitemShut
  {NoStop}%
\bibitem [{\citenamefont {{STAR
  Collaboration}}(2010)}]{star_collaboration_experimental_2010}%
  \BibitemOpen
  \bibfield  {author} {\bibinfo {author} {\bibnamefont {{STAR
  Collaboration}}},\ }\href {http://arxiv.org/abs/1007.2613} {\bibfield
  {journal} {\bibinfo  {journal} {arXiv:1007.2613 [nucl-ex]}\ } (\bibinfo
  {year} {2010})},\ \bibinfo {note} {arXiv: 1007.2613}\BibitemShut {NoStop}%
\bibitem [{\citenamefont {Heinz}\ \emph {et~al.}(2015)\citenamefont {Heinz},
  \citenamefont {Sorensen}, \citenamefont {Deshpande}, \citenamefont
  {Gagliardi}, \citenamefont {Karsch}, \citenamefont {Lappi}, \citenamefont
  {Meziani}, \citenamefont {Milner}, \citenamefont {Muller}, \citenamefont
  {Nagle}, \citenamefont {Qiu}, \citenamefont {Rajagopal}, \citenamefont
  {Roland},\ and\ \citenamefont {Venugopalan}}]{heinz_exploring_2015}%
  \BibitemOpen
  \bibfield  {author} {\bibinfo {author} {\bibfnamefont {U.}~\bibnamefont
  {Heinz}}, \bibinfo {author} {\bibfnamefont {P.}~\bibnamefont {Sorensen}},
  \bibinfo {author} {\bibfnamefont {A.}~\bibnamefont {Deshpande}}, \bibinfo
  {author} {\bibfnamefont {C.}~\bibnamefont {Gagliardi}}, \bibinfo {author}
  {\bibfnamefont {F.}~\bibnamefont {Karsch}}, \bibinfo {author} {\bibfnamefont
  {T.}~\bibnamefont {Lappi}}, \bibinfo {author} {\bibfnamefont {Z.-E.}\
  \bibnamefont {Meziani}}, \bibinfo {author} {\bibfnamefont {R.}~\bibnamefont
  {Milner}}, \bibinfo {author} {\bibfnamefont {B.}~\bibnamefont {Muller}},
  \bibinfo {author} {\bibfnamefont {J.}~\bibnamefont {Nagle}}, \bibinfo
  {author} {\bibfnamefont {J.-W.}\ \bibnamefont {Qiu}}, \bibinfo {author}
  {\bibfnamefont {K.}~\bibnamefont {Rajagopal}}, \bibinfo {author}
  {\bibfnamefont {G.}~\bibnamefont {Roland}}, \ and\ \bibinfo {author}
  {\bibfnamefont {R.}~\bibnamefont {Venugopalan}},\ }\href
  {http://arxiv.org/abs/1501.06477} {\bibfield  {journal} {\bibinfo  {journal}
  {arXiv:1501.06477 [hep-ex, physics:hep-ph, physics:nucl-ex,
  physics:nucl-th]}\ } (\bibinfo {year} {2015})},\ \bibinfo {note} {arXiv:
  1501.06477}\BibitemShut {NoStop}%
\bibitem [{\citenamefont {Deutsch}(1991)}]{deutsch_quantum_1991}%
  \BibitemOpen
  \bibfield  {author} {\bibinfo {author} {\bibfnamefont {J.~M.}\ \bibnamefont
  {Deutsch}},\ }\href {\doibase 10.1103/PhysRevA.43.2046} {\bibfield  {journal}
  {\bibinfo  {journal} {Physical Review A}\ }\textbf {\bibinfo {volume} {43}},\
  \bibinfo {pages} {2046} (\bibinfo {year} {1991})}\BibitemShut {NoStop}%
\bibitem [{\citenamefont {Srednicki}(1994)}]{srednicki_chaos_1994}%
  \BibitemOpen
  \bibfield  {author} {\bibinfo {author} {\bibfnamefont {M.}~\bibnamefont
  {Srednicki}},\ }\href {\doibase 10.1103/PhysRevE.50.888} {\bibfield
  {journal} {\bibinfo  {journal} {Physical Review E}\ }\textbf {\bibinfo
  {volume} {50}},\ \bibinfo {pages} {888} (\bibinfo {year} {1994})}\BibitemShut
  {NoStop}%
\bibitem [{\citenamefont {D'Alessio}\ \emph {et~al.}(2016)\citenamefont
  {D'Alessio}, \citenamefont {Kafri}, \citenamefont {Polkovnikov},\ and\
  \citenamefont {Rigol}}]{dalessio_quantum_2016}%
  \BibitemOpen
  \bibfield  {author} {\bibinfo {author} {\bibfnamefont {L.}~\bibnamefont
  {D'Alessio}}, \bibinfo {author} {\bibfnamefont {Y.}~\bibnamefont {Kafri}},
  \bibinfo {author} {\bibfnamefont {A.}~\bibnamefont {Polkovnikov}}, \ and\
  \bibinfo {author} {\bibfnamefont {M.}~\bibnamefont {Rigol}},\ }\href
  {\doibase 10.1080/00018732.2016.1198134} {\bibfield  {journal} {\bibinfo
  {journal} {Advances in Physics}\ }\textbf {\bibinfo {volume} {65}},\ \bibinfo
  {pages} {239} (\bibinfo {year} {2016})},\ \bibinfo {note} {arXiv:
  1509.06411}\BibitemShut {NoStop}%
\bibitem [{\citenamefont {White}\ \emph {et~al.}(2018)\citenamefont {White},
  \citenamefont {Zaletel}, \citenamefont {Mong},\ and\ \citenamefont
  {Refael}}]{white_quantum_2018}%
  \BibitemOpen
  \bibfield  {author} {\bibinfo {author} {\bibfnamefont {C.~D.}\ \bibnamefont
  {White}}, \bibinfo {author} {\bibfnamefont {M.}~\bibnamefont {Zaletel}},
  \bibinfo {author} {\bibfnamefont {R.~S.~K.}\ \bibnamefont {Mong}}, \ and\
  \bibinfo {author} {\bibfnamefont {G.}~\bibnamefont {Refael}},\ }\href
  {\doibase 10.1103/PhysRevB.97.035127} {\bibfield  {journal} {\bibinfo
  {journal} {Physical Review B}\ }\textbf {\bibinfo {volume} {97}},\ \bibinfo
  {pages} {035127} (\bibinfo {year} {2018})}\BibitemShut {NoStop}%
\bibitem [{\citenamefont {Ye}\ \emph {et~al.}(2020)\citenamefont {Ye},
  \citenamefont {Machado}, \citenamefont {White}, \citenamefont {Mong},\ and\
  \citenamefont {Yao}}]{ye_emergent_2019}%
  \BibitemOpen
  \bibfield  {author} {\bibinfo {author} {\bibfnamefont {B.}~\bibnamefont
  {Ye}}, \bibinfo {author} {\bibfnamefont {F.}~\bibnamefont {Machado}},
  \bibinfo {author} {\bibfnamefont {C.~D.}\ \bibnamefont {White}}, \bibinfo
  {author} {\bibfnamefont {R.~S.}\ \bibnamefont {Mong}}, \ and\ \bibinfo
  {author} {\bibfnamefont {N.~Y.}\ \bibnamefont {Yao}},\ }\href {\doibase
  10.1103/physrevlett.125.030601} {\bibfield  {journal} {\bibinfo  {journal}
  {Physical Review Letters}\ }\textbf {\bibinfo {volume} {125}} (\bibinfo
  {year} {2020}),\ 10.1103/physrevlett.125.030601}\BibitemShut {NoStop}%
\bibitem [{\citenamefont {Rakovszky}\ \emph {et~al.}(2022)\citenamefont
  {Rakovszky}, \citenamefont {von Keyserlingk},\ and\ \citenamefont
  {Pollmann}}]{rakovszky_dissipation-assisted_2020}%
  \BibitemOpen
  \bibfield  {author} {\bibinfo {author} {\bibfnamefont {T.}~\bibnamefont
  {Rakovszky}}, \bibinfo {author} {\bibfnamefont {C.~W.}\ \bibnamefont {von
  Keyserlingk}}, \ and\ \bibinfo {author} {\bibfnamefont {F.}~\bibnamefont
  {Pollmann}},\ }\href {\doibase 10.1103/PhysRevB.105.075131} {\bibfield
  {journal} {\bibinfo  {journal} {Phys. Rev. B}\ }\textbf {\bibinfo {volume}
  {105}},\ \bibinfo {pages} {075131} (\bibinfo {year} {2022})}\BibitemShut
  {NoStop}%
\bibitem [{\citenamefont {Ho}\ and\ \citenamefont
  {Abanin}(2017)}]{ho_entanglement_2017}%
  \BibitemOpen
  \bibfield  {author} {\bibinfo {author} {\bibfnamefont {W.~W.}\ \bibnamefont
  {Ho}}\ and\ \bibinfo {author} {\bibfnamefont {D.~A.}\ \bibnamefont
  {Abanin}},\ }\href {\doibase 10.1103/PhysRevB.95.094302} {\bibfield
  {journal} {\bibinfo  {journal} {Physical Review B}\ }\textbf {\bibinfo
  {volume} {95}},\ \bibinfo {pages} {094302} (\bibinfo {year} {2017})},\
  \bibinfo {note} {publisher: American Physical Society}\BibitemShut {NoStop}%
\bibitem [{\citenamefont {von Keyserlingk}\ \emph {et~al.}(2018)\citenamefont
  {von Keyserlingk}, \citenamefont {Rakovszky}, \citenamefont {Pollmann},\ and\
  \citenamefont {Sondhi}}]{von_keyserlingk_operator_2018}%
  \BibitemOpen
  \bibfield  {author} {\bibinfo {author} {\bibfnamefont {C.}~\bibnamefont {von
  Keyserlingk}}, \bibinfo {author} {\bibfnamefont {T.}~\bibnamefont
  {Rakovszky}}, \bibinfo {author} {\bibfnamefont {F.}~\bibnamefont {Pollmann}},
  \ and\ \bibinfo {author} {\bibfnamefont {S.}~\bibnamefont {Sondhi}},\ }\href
  {\doibase 10.1103/PhysRevX.8.021013} {\bibfield  {journal} {\bibinfo
  {journal} {Physical Review X}\ }\textbf {\bibinfo {volume} {8}},\ \bibinfo
  {pages} {021013} (\bibinfo {year} {2018})},\ \bibinfo {note} {arXiv:
  1705.08910}\BibitemShut {NoStop}%
\bibitem [{\citenamefont {Nahum}\ \emph {et~al.}(2017)\citenamefont {Nahum},
  \citenamefont {Ruhman}, \citenamefont {Vijay},\ and\ \citenamefont
  {Haah}}]{nahum_quantum_2017}%
  \BibitemOpen
  \bibfield  {author} {\bibinfo {author} {\bibfnamefont {A.}~\bibnamefont
  {Nahum}}, \bibinfo {author} {\bibfnamefont {J.}~\bibnamefont {Ruhman}},
  \bibinfo {author} {\bibfnamefont {S.}~\bibnamefont {Vijay}}, \ and\ \bibinfo
  {author} {\bibfnamefont {J.}~\bibnamefont {Haah}},\ }\href {\doibase
  10.1103/PhysRevX.7.031016} {\bibfield  {journal} {\bibinfo  {journal}
  {Physical Review X}\ }\textbf {\bibinfo {volume} {7}},\ \bibinfo {pages}
  {031016} (\bibinfo {year} {2017})},\ \bibinfo {note} {publisher: American
  Physical Society}\BibitemShut {NoStop}%
\bibitem [{\citenamefont {Khemani}\ \emph {et~al.}(2018)\citenamefont
  {Khemani}, \citenamefont {Vishwanath},\ and\ \citenamefont
  {Huse}}]{khemani_operator_2018}%
  \BibitemOpen
  \bibfield  {author} {\bibinfo {author} {\bibfnamefont {V.}~\bibnamefont
  {Khemani}}, \bibinfo {author} {\bibfnamefont {A.}~\bibnamefont {Vishwanath}},
  \ and\ \bibinfo {author} {\bibfnamefont {D.~A.}\ \bibnamefont {Huse}},\
  }\href {\doibase 10.1103/PhysRevX.8.031057} {\bibfield  {journal} {\bibinfo
  {journal} {Physical Review X}\ }\textbf {\bibinfo {volume} {8}},\ \bibinfo
  {pages} {031057} (\bibinfo {year} {2018})}\BibitemShut {NoStop}%
\bibitem [{\citenamefont {Rakovszky}\ \emph {et~al.}(2018)\citenamefont
  {Rakovszky}, \citenamefont {Pollmann},\ and\ \citenamefont {von
  Keyserlingk}}]{rakovszky_diffusive_2017}%
  \BibitemOpen
  \bibfield  {author} {\bibinfo {author} {\bibfnamefont {T.}~\bibnamefont
  {Rakovszky}}, \bibinfo {author} {\bibfnamefont {F.}~\bibnamefont {Pollmann}},
  \ and\ \bibinfo {author} {\bibfnamefont {C.~W.}\ \bibnamefont {von
  Keyserlingk}},\ }\href {\doibase 10.1103/PhysRevX.8.031058} {\bibfield
  {journal} {\bibinfo  {journal} {Phys. Rev. X}\ }\textbf {\bibinfo {volume}
  {8}},\ \bibinfo {pages} {031058} (\bibinfo {year} {2018})}\BibitemShut
  {NoStop}%
\bibitem [{\citenamefont {Kvorning}\ \emph {et~al.}(2022)\citenamefont
  {Kvorning}, \citenamefont {Herviou},\ and\ \citenamefont
  {Bardarson}}]{kvorning_time-evolution_2021}%
  \BibitemOpen
  \bibfield  {author} {\bibinfo {author} {\bibfnamefont {T.~K.}\ \bibnamefont
  {Kvorning}}, \bibinfo {author} {\bibfnamefont {L.}~\bibnamefont {Herviou}}, \
  and\ \bibinfo {author} {\bibfnamefont {J.~H.}\ \bibnamefont {Bardarson}},\
  }\href {\doibase 10.21468/SciPostPhys.13.4.080} {\bibfield  {journal}
  {\bibinfo  {journal} {SciPost Phys.}\ }\textbf {\bibinfo {volume} {13}},\
  \bibinfo {pages} {080} (\bibinfo {year} {2022})}\BibitemShut {NoStop}%
\bibitem [{\citenamefont {Altman}(2018)}]{altman_ehud_computing_2018}%
  \BibitemOpen
  \bibfield  {author} {\bibinfo {author} {\bibfnamefont {E.}~\bibnamefont
  {Altman}},\ }\href {https://online.kitp.ucsb.edu/online/dynq_c18/altman/}
  {\enquote {\bibinfo {title} {Computing quantum thermalization dynamics},}\ }
  (\bibinfo {year} {2018})\BibitemShut {NoStop}%
\bibitem [{\citenamefont {Parker}\ \emph {et~al.}(2019)\citenamefont {Parker},
  \citenamefont {Cao}, \citenamefont {Avdoshkin}, \citenamefont {Scaffidi},\
  and\ \citenamefont {Altman}}]{parker_universal_2019}%
  \BibitemOpen
  \bibfield  {author} {\bibinfo {author} {\bibfnamefont {D.~E.}\ \bibnamefont
  {Parker}}, \bibinfo {author} {\bibfnamefont {X.}~\bibnamefont {Cao}},
  \bibinfo {author} {\bibfnamefont {A.}~\bibnamefont {Avdoshkin}}, \bibinfo
  {author} {\bibfnamefont {T.}~\bibnamefont {Scaffidi}}, \ and\ \bibinfo
  {author} {\bibfnamefont {E.}~\bibnamefont {Altman}},\ }\href {\doibase
  10.1103/PhysRevX.9.041017} {\bibfield  {journal} {\bibinfo  {journal}
  {Physical Review X}\ }\textbf {\bibinfo {volume} {9}},\ \bibinfo {pages}
  {041017} (\bibinfo {year} {2019})},\ \bibinfo {note} {arXiv:
  1812.08657}\BibitemShut {NoStop}%
\bibitem [{\citenamefont {Cao}(2021)}]{cao_statistical_2021}%
  \BibitemOpen
  \bibfield  {author} {\bibinfo {author} {\bibfnamefont {X.}~\bibnamefont
  {Cao}},\ }\href {\doibase 10.1088/1751-8121/abe77c} {\bibfield  {journal}
  {\bibinfo  {journal} {Journal of Physics A: Mathematical and Theoretical}\
  }\textbf {\bibinfo {volume} {54}},\ \bibinfo {pages} {144001} (\bibinfo
  {year} {2021})},\ \bibinfo {note} {arXiv: 2012.06544}\BibitemShut {NoStop}%
\bibitem [{\citenamefont {Kim}\ and\ \citenamefont
  {Huse}(2013)}]{kim_ballistic_2013}%
  \BibitemOpen
  \bibfield  {author} {\bibinfo {author} {\bibfnamefont {H.}~\bibnamefont
  {Kim}}\ and\ \bibinfo {author} {\bibfnamefont {D.~A.}\ \bibnamefont {Huse}},\
  }\href {\doibase 10.1103/PhysRevLett.111.127205} {\bibfield  {journal}
  {\bibinfo  {journal} {Physical Review Letters}\ }\textbf {\bibinfo {volume}
  {111}},\ \bibinfo {pages} {127205} (\bibinfo {year} {2013})},\ \bibinfo
  {note} {publisher: American Physical Society}\BibitemShut {NoStop}%
\bibitem [{\citenamefont {Kim}\ \emph {et~al.}(2014)\citenamefont {Kim},
  \citenamefont {Ikeda},\ and\ \citenamefont {Huse}}]{kim_testing_2014}%
  \BibitemOpen
  \bibfield  {author} {\bibinfo {author} {\bibfnamefont {H.}~\bibnamefont
  {Kim}}, \bibinfo {author} {\bibfnamefont {T.~N.}\ \bibnamefont {Ikeda}}, \
  and\ \bibinfo {author} {\bibfnamefont {D.~A.}\ \bibnamefont {Huse}},\ }\href
  {\doibase 10.1103/PhysRevE.90.052105} {\bibfield  {journal} {\bibinfo
  {journal} {Physical Review E}\ }\textbf {\bibinfo {volume} {90}},\ \bibinfo
  {pages} {052105} (\bibinfo {year} {2014})},\ \bibinfo {note} {publisher:
  American Physical Society}\BibitemShut {NoStop}%
\bibitem [{\citenamefont {Abou-Chacra}\ \emph {et~al.}(1973)\citenamefont
  {Abou-Chacra}, \citenamefont {Thouless},\ and\ \citenamefont
  {Anderson}}]{abou-chacra_selfconsistent}%
  \BibitemOpen
  \bibfield  {author} {\bibinfo {author} {\bibfnamefont {R.}~\bibnamefont
  {Abou-Chacra}}, \bibinfo {author} {\bibfnamefont {D.~J.}\ \bibnamefont
  {Thouless}}, \ and\ \bibinfo {author} {\bibfnamefont {P.~W.}\ \bibnamefont
  {Anderson}},\ }\href {\doibase 10.1088/0022-3719/6/10/009} {\bibfield
  {journal} {\bibinfo  {journal} {Journal of Physics C: Solid State Physics}\
  }\textbf {\bibinfo {volume} {6}},\ \bibinfo {pages} {1734} (\bibinfo {year}
  {1973})}\BibitemShut {NoStop}%
\bibitem [{\citenamefont {Feenberg}(1948)}]{feenberg_note_1948}%
  \BibitemOpen
  \bibfield  {author} {\bibinfo {author} {\bibfnamefont {E.}~\bibnamefont
  {Feenberg}},\ }\href {\doibase 10.1103/PhysRev.74.206} {\bibfield  {journal}
  {\bibinfo  {journal} {Physical Review}\ }\textbf {\bibinfo {volume} {74}},\
  \bibinfo {pages} {206} (\bibinfo {year} {1948})}\BibitemShut {NoStop}%
\bibitem [{\citenamefont {Watson}(1957)}]{watson_multiple_1957}%
  \BibitemOpen
  \bibfield  {author} {\bibinfo {author} {\bibfnamefont {K.~M.}\ \bibnamefont
  {Watson}},\ }\href {\doibase 10.1103/PhysRev.105.1388} {\bibfield  {journal}
  {\bibinfo  {journal} {Physical Review}\ }\textbf {\bibinfo {volume} {105}},\
  \bibinfo {pages} {1388} (\bibinfo {year} {1957})},\ \bibinfo {note}
  {publisher: American Physical Society}\BibitemShut {NoStop}%
\bibitem [{\citenamefont {von Keyserlingk}\ \emph {et~al.}(2022)\citenamefont
  {von Keyserlingk}, \citenamefont {Pollmann},\ and\ \citenamefont
  {Rakovszky}}]{vonkeyserlingkOperatorBackflowClassical2021}%
  \BibitemOpen
  \bibfield  {author} {\bibinfo {author} {\bibfnamefont {C.}~\bibnamefont {von
  Keyserlingk}}, \bibinfo {author} {\bibfnamefont {F.}~\bibnamefont
  {Pollmann}}, \ and\ \bibinfo {author} {\bibfnamefont {T.}~\bibnamefont
  {Rakovszky}},\ }\href {\doibase 10.1103/PhysRevB.105.245101} {\bibfield
  {journal} {\bibinfo  {journal} {Phys. Rev. B}\ }\textbf {\bibinfo {volume}
  {105}},\ \bibinfo {pages} {245101} (\bibinfo {year} {2022})}\BibitemShut
  {NoStop}%
\bibitem [{\citenamefont {Zwanzig}(1961)}]{zwanzig_memory_1961}%
  \BibitemOpen
  \bibfield  {author} {\bibinfo {author} {\bibfnamefont {R.}~\bibnamefont
  {Zwanzig}},\ }\href {\doibase 10.1103/PhysRev.124.983} {\bibfield  {journal}
  {\bibinfo  {journal} {Physical Review}\ }\textbf {\bibinfo {volume} {124}},\
  \bibinfo {pages} {983} (\bibinfo {year} {1961})},\ \bibinfo {note}
  {publisher: American Physical Society}\BibitemShut {NoStop}%
\bibitem [{\citenamefont {Mori}(1965)}]{mori_transport_1965}%
  \BibitemOpen
  \bibfield  {author} {\bibinfo {author} {\bibfnamefont {H.}~\bibnamefont
  {Mori}},\ }\href {\doibase 10.1143/PTP.33.423} {\bibfield  {journal}
  {\bibinfo  {journal} {Progress of Theoretical Physics}\ }\textbf {\bibinfo
  {volume} {33}},\ \bibinfo {pages} {423} (\bibinfo {year} {1965})}\BibitemShut
  {NoStop}%
\bibitem [{\citenamefont {Grabert}(1982)}]{grabert_projection_1982}%
  \BibitemOpen
  \bibfield  {author} {\bibinfo {author} {\bibfnamefont {H.}~\bibnamefont
  {Grabert}},\ }\href@noop {} {\emph {\bibinfo {title} {Projection {Operator}
  {Techniques} in {Nonequilibrium} {Statistical} {Mechanics}}}},\ \bibinfo
  {series} {Springer {Tracts} in {Modern}}\ No.~\bibinfo {number} {95}\
  (\bibinfo  {publisher} {Springer-Verlag},\ \bibinfo {address} {Berlin},\
  \bibinfo {year} {1982})\ \bibinfo {note} {section: 164 str.}\BibitemShut
  {Stop}%
\bibitem [{\citenamefont {Viswanath}\ and\ \citenamefont
  {Müller}(1994)}]{viswanath_recursion_1994}%
  \BibitemOpen
  \bibfield  {author} {\bibinfo {author} {\bibfnamefont {V.~S.}\ \bibnamefont
  {Viswanath}}\ and\ \bibinfo {author} {\bibfnamefont {G.}~\bibnamefont
  {Müller}},\ }\href@noop {} {\emph {\bibinfo {title} {The Recursion Method:
  Application to Many-Body Dynamics}}},\ \bibinfo {series} {Lecture notes in
  physics}\ No.\ \bibinfo {number} {m23}\ (\bibinfo  {publisher}
  {Springer-Verlag},\ \bibinfo {address} {Berlin ; New York},\ \bibinfo {year}
  {1994})\BibitemShut {NoStop}%
\bibitem [{Note4()}]{Note4}%
  \BibitemOpen
  \bibinfo {note} {I am a priori skeptical that this convergence is
  trustworthy. The authors of \protect \rev@citealpnum
  {viswanath_recursion_1994} are likewise skeptical, and in Sec.~8.1 they argue
  that the pole structure of a response function depends on all the
  continued-fraction coefficients (hence all the moments $M_n$). But the
  numerics of \protect \rev@citealpnum {viswanath_recursion_1994} Sec.~8.1 are
  marred by the fact they work at finite system size, rather than working
  directly in Fourier space as in \protect \rev@citealpnum
  {jung_transport_2006}. \protect \rev@citealpnum {jung_transport_2006}
  performs simple truncation at order $n \sim 20$, but then extrapolates the
  resulting DC conductivity to $n = \infty $. This is no worse than most other
  numerical methods, which likewise rely on convergence testing, and the result
  of \protect \rev@citealpnum {jung_transport_2006}---that certain
  $SU(2)$-symmetric perturbations to the Heisenberg model do \protect \textit
  {not} break anomalous diffusion---is consistent with very different recent
  work \cite {de_nardis_stability_2021}. I am curious (1) whether the
  extrapolation can be physically justified, and (2) whether it gives correct
  diffusion coefficients in non-integrable systems}\BibitemShut {NoStop}%
\bibitem [{\citenamefont {Vanhecke}\ \emph {et~al.}(2021)\citenamefont
  {Vanhecke}, \citenamefont {Devoogdt}, \citenamefont {Verstraete},\ and\
  \citenamefont {Vanderstraeten}}]{vanhecke_simulating_2021}%
  \BibitemOpen
  \bibfield  {author} {\bibinfo {author} {\bibfnamefont {B.}~\bibnamefont
  {Vanhecke}}, \bibinfo {author} {\bibfnamefont {D.}~\bibnamefont {Devoogdt}},
  \bibinfo {author} {\bibfnamefont {F.}~\bibnamefont {Verstraete}}, \ and\
  \bibinfo {author} {\bibfnamefont {L.}~\bibnamefont {Vanderstraeten}},\ }\href
  {http://arxiv.org/abs/2112.01507} {\bibfield  {journal} {\bibinfo  {journal}
  {arXiv:2112.01507 [cond-mat, physics:quant-ph]}\ } (\bibinfo {year}
  {2021})},\ \bibinfo {note} {arXiv: 2112.01507}\BibitemShut {NoStop}%
\bibitem [{\citenamefont {Kovtun}\ and\ \citenamefont
  {Yaffe}(2003)}]{kovtun_hydrodynamic_2003}%
  \BibitemOpen
  \bibfield  {author} {\bibinfo {author} {\bibfnamefont {P.}~\bibnamefont
  {Kovtun}}\ and\ \bibinfo {author} {\bibfnamefont {L.~G.}\ \bibnamefont
  {Yaffe}},\ }\href {\doibase 10.1103/PhysRevD.68.025007} {\bibfield  {journal}
  {\bibinfo  {journal} {Physical Review D}\ }\textbf {\bibinfo {volume} {68}},\
  \bibinfo {pages} {025007} (\bibinfo {year} {2003})},\ \bibinfo {note} {arXiv:
  hep-th/0303010}\BibitemShut {NoStop}%
\bibitem [{\citenamefont {Mukerjee}\ \emph {et~al.}(2006)\citenamefont
  {Mukerjee}, \citenamefont {Oganesyan},\ and\ \citenamefont
  {Huse}}]{mukerjee_towards_2006}%
  \BibitemOpen
  \bibfield  {author} {\bibinfo {author} {\bibfnamefont {S.}~\bibnamefont
  {Mukerjee}}, \bibinfo {author} {\bibfnamefont {V.}~\bibnamefont {Oganesyan}},
  \ and\ \bibinfo {author} {\bibfnamefont {D.}~\bibnamefont {Huse}},\ }\href
  {\doibase 10.1103/PhysRevB.73.035113} {\bibfield  {journal} {\bibinfo
  {journal} {Physical Review B}\ }\textbf {\bibinfo {volume} {73}} (\bibinfo
  {year} {2006}),\ 10.1103/PhysRevB.73.035113},\ \bibinfo {note} {arXiv:
  cond-mat/0503177}\BibitemShut {NoStop}%
\bibitem [{\citenamefont {Spohn}(2014)}]{spohn_nonlinear_2014}%
  \BibitemOpen
  \bibfield  {author} {\bibinfo {author} {\bibfnamefont {H.}~\bibnamefont
  {Spohn}},\ }\href {\doibase 10.1007/s10955-014-0933-y} {\bibfield  {journal}
  {\bibinfo  {journal} {Journal of Statistical Physics}\ }\textbf {\bibinfo
  {volume} {154}},\ \bibinfo {pages} {1191} (\bibinfo {year} {2014})},\
  \bibinfo {note} {arXiv: 1305.6412}\BibitemShut {NoStop}%
\bibitem [{\citenamefont {Lux}\ \emph {et~al.}(2014)\citenamefont {Lux},
  \citenamefont {Müller}, \citenamefont {Mitra},\ and\ \citenamefont
  {Rosch}}]{lux_hydrodynamic_2014}%
  \BibitemOpen
  \bibfield  {author} {\bibinfo {author} {\bibfnamefont {J.}~\bibnamefont
  {Lux}}, \bibinfo {author} {\bibfnamefont {J.}~\bibnamefont {Müller}},
  \bibinfo {author} {\bibfnamefont {A.}~\bibnamefont {Mitra}}, \ and\ \bibinfo
  {author} {\bibfnamefont {A.}~\bibnamefont {Rosch}},\ }\href {\doibase
  10.1103/PhysRevA.89.053608} {\bibfield  {journal} {\bibinfo  {journal}
  {Physical Review A}\ }\textbf {\bibinfo {volume} {89}},\ \bibinfo {pages}
  {053608} (\bibinfo {year} {2014})},\ \bibinfo {note} {arXiv:
  1311.7644}\BibitemShut {NoStop}%
\bibitem [{\citenamefont {Kirkpatrick}\ \emph {et~al.}(2021)\citenamefont
  {Kirkpatrick}, \citenamefont {Belitz},\ and\ \citenamefont
  {Dorfman}}]{kirkpatrick_non-hydrodynamic_2021}%
  \BibitemOpen
  \bibfield  {author} {\bibinfo {author} {\bibfnamefont {T.~R.}\ \bibnamefont
  {Kirkpatrick}}, \bibinfo {author} {\bibfnamefont {D.}~\bibnamefont {Belitz}},
  \ and\ \bibinfo {author} {\bibfnamefont {J.~R.}\ \bibnamefont {Dorfman}},\
  }\href {\doibase 10.1103/PhysRevE.104.024111} {\bibfield  {journal} {\bibinfo
   {journal} {Phys. Rev. E}\ }\textbf {\bibinfo {volume} {104}},\ \bibinfo
  {pages} {024111} (\bibinfo {year} {2021})}\BibitemShut {NoStop}%
\bibitem [{\citenamefont {Jung}\ \emph {et~al.}(2006)\citenamefont {Jung},
  \citenamefont {Helmes},\ and\ \citenamefont {Rosch}}]{jung_transport_2006}%
  \BibitemOpen
  \bibfield  {author} {\bibinfo {author} {\bibfnamefont {P.}~\bibnamefont
  {Jung}}, \bibinfo {author} {\bibfnamefont {R.~W.}\ \bibnamefont {Helmes}}, \
  and\ \bibinfo {author} {\bibfnamefont {A.}~\bibnamefont {Rosch}},\ }\href
  {\doibase 10.1103/PhysRevLett.96.067202} {\bibfield  {journal} {\bibinfo
  {journal} {Physical Review Letters}\ }\textbf {\bibinfo {volume} {96}},\
  \bibinfo {pages} {067202} (\bibinfo {year} {2006})},\ \bibinfo {note} {arXiv:
  cond-mat/0509615}\BibitemShut {NoStop}%
\bibitem [{\citenamefont {De~Nardis}\ \emph {et~al.}(2021)\citenamefont
  {De~Nardis}, \citenamefont {Gopalakrishnan}, \citenamefont {Vasseur},\ and\
  \citenamefont {Ware}}]{de_nardis_stability_2021}%
  \BibitemOpen
  \bibfield  {author} {\bibinfo {author} {\bibfnamefont {J.}~\bibnamefont
  {De~Nardis}}, \bibinfo {author} {\bibfnamefont {S.}~\bibnamefont
  {Gopalakrishnan}}, \bibinfo {author} {\bibfnamefont {R.}~\bibnamefont
  {Vasseur}}, \ and\ \bibinfo {author} {\bibfnamefont {B.}~\bibnamefont
  {Ware}},\ }\href {\doibase 10.1103/PhysRevLett.127.057201} {\bibfield
  {journal} {\bibinfo  {journal} {Physical Review Letters}\ }\textbf {\bibinfo
  {volume} {127}},\ \bibinfo {pages} {057201} (\bibinfo {year} {2021})},\
  \bibinfo {note} {arXiv: 2102.02219}\BibitemShut {NoStop}%
\bibitem [{\citenamefont {Haegeman}(2021)}]{jutho_juthokrylovkitjl_2021}%
  \BibitemOpen
  \bibfield  {author} {\bibinfo {author} {\bibfnamefont {J.}~\bibnamefont
  {Haegeman}},\ }\href {https://github.com/Jutho/KrylovKit.jl} {\enquote
  {\bibinfo {title} {Jutho/{KrylovKit}.jl},}\ } (\bibinfo {year} {2021}),\
  \bibinfo {note} {original-date: 2017-09-19T10:25:33Z}\BibitemShut {NoStop}%
\bibitem [{\citenamefont {Drabold}\ and\ \citenamefont
  {Sankey}(1993)}]{drabold_maximum_1993}%
  \BibitemOpen
  \bibfield  {author} {\bibinfo {author} {\bibfnamefont {D.~A.}\ \bibnamefont
  {Drabold}}\ and\ \bibinfo {author} {\bibfnamefont {O.~F.}\ \bibnamefont
  {Sankey}},\ }\href {\doibase 10.1103/PhysRevLett.70.3631} {\bibfield
  {journal} {\bibinfo  {journal} {Physical Review Letters}\ }\textbf {\bibinfo
  {volume} {70}},\ \bibinfo {pages} {3631} (\bibinfo {year}
  {1993})}\BibitemShut {NoStop}%
\bibitem [{\citenamefont {Silver}\ and\ \citenamefont
  {Röder}(1994)}]{silver_densities_1994}%
  \BibitemOpen
  \bibfield  {author} {\bibinfo {author} {\bibfnamefont {R.}~\bibnamefont
  {Silver}}\ and\ \bibinfo {author} {\bibfnamefont {H.}~\bibnamefont
  {Röder}},\ }\href {\doibase 10.1142/S0129183194000842} {\bibfield  {journal}
  {\bibinfo  {journal} {International Journal of Modern Physics C}\ }\textbf
  {\bibinfo {volume} {05}},\ \bibinfo {pages} {735} (\bibinfo {year} {1994})},\
  \bibinfo {note} {publisher: World Scientific Publishing Co.}\BibitemShut
  {Stop}%
\bibitem [{\citenamefont {Silver}\ and\ \citenamefont
  {Röder}(1997)}]{silver_calculation_1997}%
  \BibitemOpen
  \bibfield  {author} {\bibinfo {author} {\bibfnamefont {R.~N.}\ \bibnamefont
  {Silver}}\ and\ \bibinfo {author} {\bibfnamefont {H.}~\bibnamefont
  {Röder}},\ }\href {\doibase 10.1103/PhysRevE.56.4822} {\bibfield  {journal}
  {\bibinfo  {journal} {Physical Review E}\ }\textbf {\bibinfo {volume} {56}},\
  \bibinfo {pages} {4822} (\bibinfo {year} {1997})}\BibitemShut {NoStop}%
\bibitem [{\citenamefont {Weisse}\ \emph {et~al.}(2006)\citenamefont {Weisse},
  \citenamefont {Wellein}, \citenamefont {Alvermann},\ and\ \citenamefont
  {Fehske}}]{weisse_kernel_2006}%
  \BibitemOpen
  \bibfield  {author} {\bibinfo {author} {\bibfnamefont {A.}~\bibnamefont
  {Weisse}}, \bibinfo {author} {\bibfnamefont {G.}~\bibnamefont {Wellein}},
  \bibinfo {author} {\bibfnamefont {A.}~\bibnamefont {Alvermann}}, \ and\
  \bibinfo {author} {\bibfnamefont {H.}~\bibnamefont {Fehske}},\ }\href
  {\doibase 10.1103/RevModPhys.78.275} {\bibfield  {journal} {\bibinfo
  {journal} {Reviews of Modern Physics}\ }\textbf {\bibinfo {volume} {78}},\
  \bibinfo {pages} {275} (\bibinfo {year} {2006})},\ \bibinfo {note} {arXiv:
  cond-mat/0504627}\BibitemShut {NoStop}%
\end{thebibliography}%

\appendix

\section{Particle loss in the operator graph picture} \label{ss:fact-loss}

\begin{figure}
  \includegraphics[width=0.45\textwidth]{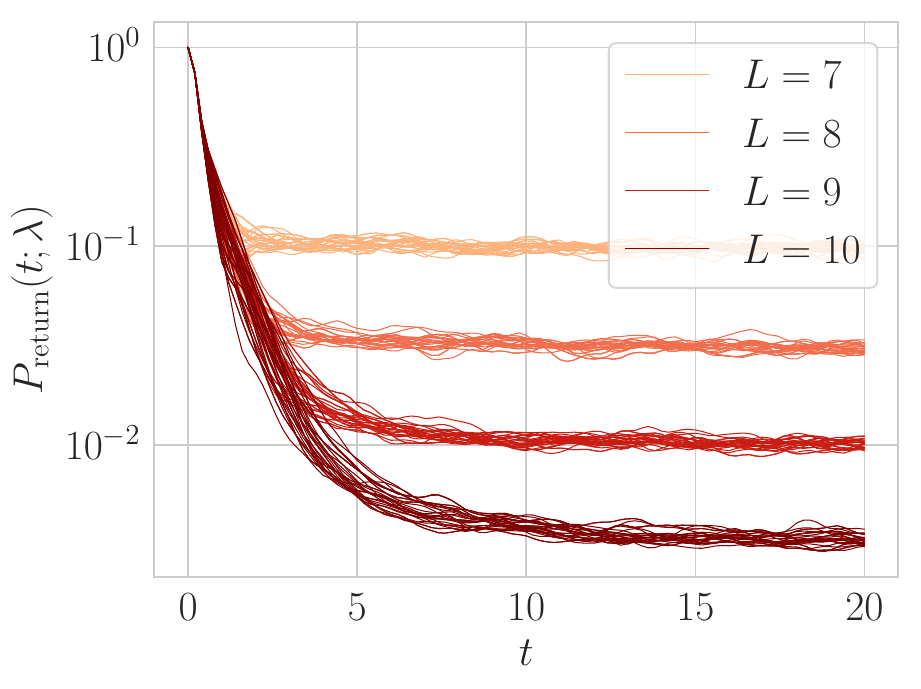}
  \caption{\textbf{Probability of return to the set of operators of diameter $\le 4$} starting from a Pauli string $\sigma^{\bm \lambda}$ of diameter $4$ in the modified dynamics of Eq.~\eqref{eq:Hgraph-ret-prob}.
    Colors show system size $7 \le L \le 11$.
    Individual lines show different (randomly chosen) initial operators.
    In the large-system limit, the particle will not return;
    for finite system size,
    the long-time value is set by the Hilbert space dimension.
  }
  \label{fig:ret-prob}
\end{figure}

To understand why $\pool_l$ causes particle loss,
consider some Pauli string $\sigma^{\bm \lambda} \in \pool_{l-1}$.
Imagine a particle on that string hops to a string $\sigma^{\bm \mu} \in \pool_{l}$.
Let us compute at leading order in $tJ$ the probability that it returns to $\pool_{l-1}$.
Recall
\begin{align}
  \begin{split}
    \mathcal M_{l'}(\bm \lambda)
    &= \{
    \bm \nu : \text{$\diam \bm \nu = 0$ and $L_{\bm \nu\bm \lambda} \ne 0$}
    \}\;;
  \end{split}
\end{align}
the nontrivial cases are $l-1$, $l$, and $l+1$.
In Sec.~\ref{ss:graph-structure} I implicitly argued that
\begin{align}
 \frac{ |\mathcal M_{l-1}(\bm \lambda)|}{|\mathcal M_{l}(\bm \lambda)|} \propto l^{-1}\;.
\end{align}
At time $t$ the state is (at leading order in $t$)
\begin{align}
  \ket \psi = \ket{\bm\lambda} - it \sum_{\bm \nu} L_{\bm\nu\bm\lambda}\ket{\bm \nu}\;.
\end{align}
So the probability that the particle hops to sites in $\pool_{l-1}$ and $\pool_{l}$ respectively are
\begin{align}
  P_{l-1} &= \sum_{\bm \nu \in \mathcal M_{l-1}(\bm \lambda)} |L_{\bm \nu\bm\lambda} |^2 \\
  P_{l} &= \sum_{\bm \mu \in \mathcal M_{l}(\bm \lambda)} |L_{\bm \mu\bm\lambda} |^2\;.
\end{align}
Since the matrix elements of the Liouvillian are $L_{\bm\mu\bm\nu} \propto J_{\alpha\beta}$ and we take all the $J_{\alpha\beta}$ of comparable magnitude,
\begin{align}
  \label{eq:prob-ratio}
  \frac {P_{l-1}}{P_l} \approx \frac{ |\mathcal M_{l-1}(\bm \lambda)|}{|\mathcal M_{l}(\bm \lambda)|} \propto l^{-1}\;.
\end{align}
To phrase this in terms of amplitudes rather than probabilities,
the propagator to other sites in $\pool_{l-1}$ is suppressed by a factor $l^{-1/2}$.

Eq.~\ref{eq:prob-ratio} is only well-justified for short times.
For intermediate times hopping from $\pool_{l}$ to $\pool_{l+1}$ becomes important.
Heuristically, the fact that
\begin{align}
  \frac{n_-}{n_+}
  \equiv
  \frac {\overline{|\mathcal M_{l-1}(\lambda) |}}
  {\overline{|\mathcal M_{l+1}(\lambda) |}}
  = \alpha,\ \alpha > 1
\end{align}
means that the particle, having bounced around in $\mathcal G_l$,
is more likely to end up in $\mathcal G_{l+1}$,
or subgraphs of even larger diameter,
than in $\mathcal G_{l-1}$
(For the Ising model, I found $\alpha = 2$.)
One could make this argument more carefully by proceeding by induction.
If $\mathcal G_{l+1}$ perfectly absorbs particles (at order $l^{-1}$),
then $\mathcal G_l$ does as well, by arguments like those leading to \eqref{eq:prob-ratio};
one would continue in this way as long as $l \gg 1$.

Rather than making this inductive argument, I resort to numerics.
I check that particles which hop from $\pool_l$ to $\pool_{l+1}$ do not return
by measuring certain carefully constructed return probabilities.
Choose initial operator $\bm \lambda$ with length $l-1$.
We wish to probe whether a particle that has hopped from $\bm \lambda$ to $\pool_{l}$ will ever return, so
modify the graph Hamiltonian $H_\graph$ to delete all the connections between $\bm \lambda$ and other operators with diameter $l-1$ or $l-2$.
The result is a Hamiltonian
\begin{align}
  \label{eq:Hgraph-ret-prob}
  H_\graph' =  \sum_{\bm \mu \bm \nu} L_{\bm \mu \bm \nu}' c^\dagger_{\bm \mu} c_{\bm \nu}
\end{align}
with
\begin{align}
  L'_{\bm\lambda \bm \nu} = L'_{\bm \nu \bm \lambda} = 0, \quad \bm \nu \in \pool_{l-2} \cup \pool_{l-1}\;.
\end{align}
Then measure the probability
\begin{align}
  P_{\text{return}}(t;\bm \lambda) = \sum_{\diam \bm \mu \le l} \left|\bra{\bm \mu} e^{-iH_\graph't} \ket{\bm \lambda}\right|^2
\end{align}
that the particle,
having started on site $\bm \lambda$ in this modified graph,
has returned to an operator of diameter $d < l_{\max}$ at time $t$.
For the sake of uniformity in the connection between $\bm \lambda$ and $\pool_{l+1}$
I take $\bm \lambda$ that have nontrivial connections to $\pool_{L+1}$ on both ends---that is, $\supp \bm \lambda \subseteq \{1, \dots, l\}$ and $\sigma^{\lambda_1}, \sigma^{\lambda_l} \ne \sigma^z$.

Fig.~\ref{fig:ret-prob} shows the result of this calculation
for all $l = 4$ Pauli strings
on systems of varying length $7 \le L \le 11$.
We see that after an initial drop, the probability $P_{\text{return}}$
that the particle has returned to $\bigcup_{l' \le l} \pool_{l'}$
is exponentially small in system size.
To understand this,
note the single-particle Hamiltonian $H_\graph'$ is presumptively chaotic.
So after some time controlled by the diameter of the graph, which is $\sim L$,
the particle should be equally likely to be anywhere in the graph.
Since $\bigcup_{l' \le l} \pool_{l'}$ is an exponentially small fraction of the Hilbert space,
$P_{\text{return}}$ should be exponentially small---as indeed it is.

\section{Details of two-site toy model}\label{app:toy}

The quick calculation of the behavior of the two-site toy model in Sec.~\ref{ss:loss-toy} involved a certain amount of sleight of hand.
In this section I give a somewhat more explicit calculation.

\begin{figure}[t!]
  \includegraphics[width=0.20\textwidth]{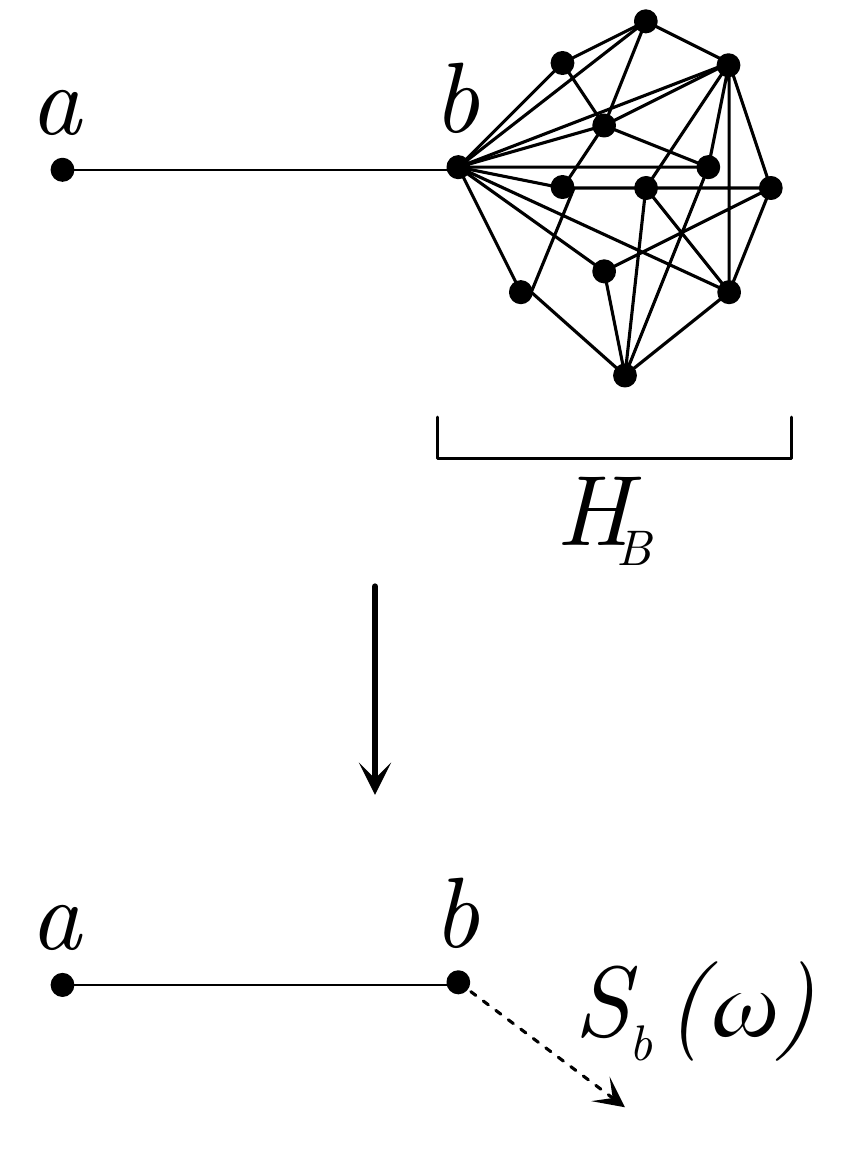}
  \caption{\textbf{Toy model} of Eq.\eqref{eq:toy-app}}
  \label{fig:toy}
\end{figure}

Consider a single-particle Hamiltonian
\begin{align}
  \label{eq:toy-app}
  H = V c^\dagger_a c_b + h.c. + H_{B}
\end{align}
where $H_B$ consists of single-particle dynamics on site $b$ together with many other sites, as in Fig.~\ref{fig:toy}.
Encapsulate the effect of $H_B$ as a self-energy $S_b$ for site $b$, considering only $H_B$:
\begin{align}
  \frac 1 {\omega - S_b(\omega)} = G_{bb;B}(\omega) \equiv \bra b \frac 1 {\omega - H_B} \ket b\;.
\end{align}
Now look at the effect of $H_B$ on site $a$.
Consider times short compared with the hopping $V$ between $a$ and $b$,
i.e. leading (second) order in $V/\omega$,
but long compared with the dynamics of $H_B$.
Then
\begin{align}
  \begin{split}
  G_{aa}(\omega) &\equiv \bra a [\omega - H]^{-1} \ket a \\
  &= \omega^{-1}  [1 + V^2 \bra b (1 + H_B/\omega + H_B^2/\omega^2 + \dots) \ket b]\\
  &\qquad\qquad+ O(V^2) \\
                 &= \omega^{-1} \left[1 + \frac {V^2/\omega} {\omega - S_b(\omega)}\right] + O(V^2/\omega^2)\;.
               \end{split}
\end{align}
But the self-energy of $a$ is
\begin{align}
  S_a(\omega) &= \omega - G_{aa}(\omega) \\
              &= \omega - \omega\left[1 -  \frac {V^2/\omega} {\omega - S_b(\omega)}\right] + O(V^2/\omega^2)\\
              &= \frac {V^2}{\omega - S_b(\omega)} + O(V^2/\omega^2)
\end{align}
Since we consider times long compared with the dynamics of $S_b$, we can take
\begin{align}
  S_b &= -i\gamma + \alpha \omega + \frac 1 2 \tau \omega^2 + \dots
\end{align}
for
\begin{align}
  S_a(\omega) = -i V^2/\gamma + O(V^2/\omega^2) + O(\omega/\gamma)
\end{align}
as in the main text.

\section{Simulating the full dynamics}\label{app:full}

Benchmarking as in Fig.~\ref{fig:diffusion} requires simulations of the full dynamics of the model \eqref{eq:tfim}.
I compute
\[ \tr [\epsilon_{k'} \varepsilon_k(t)] \]
in the full dynamics by computing the average
\begin{align}
  \tr [\epsilon_{k'} \varepsilon_k(t)]  \approx \frac 1 {N_r} \sum_{\alpha = 1}^{N_r} \bra {x_\alpha} \varepsilon_{k'} e^{iHt}\varepsilon_k e^{-iHt} \ket{x_\alpha}
\end{align}
over vectors $\ket{x_\alpha}$ with random complex normal components,
using Krylov evolution implemented in \cite{jutho_juthokrylovkitjl_2021}.
This is essentially the stochastic trace familiar from the kernel polynomial method
\cite{drabold_maximum_1993,silver_densities_1994,silver_calculation_1997,weisse_kernel_2006}.
A more sophisticated Chebyshev expansion method would presumably be more efficient,
but would not overcome the exponential scaling of the Hilbert space.
Statistical errors from sampling are small compared with methodological errors,
about which see \eqref{app:extract}.

\section{Extracting diffusion coefficients}\label{app:extract}

In this appendix I argue that large system sizes $L \gtrsim 20$
offer a window of diffusive dynamics between
short-time non-diffusive behavior
and the Thouless time.

\begin{figure}[t]
  \begin{minipage}{0.45\textwidth}
    \includegraphics[width=\textwidth]{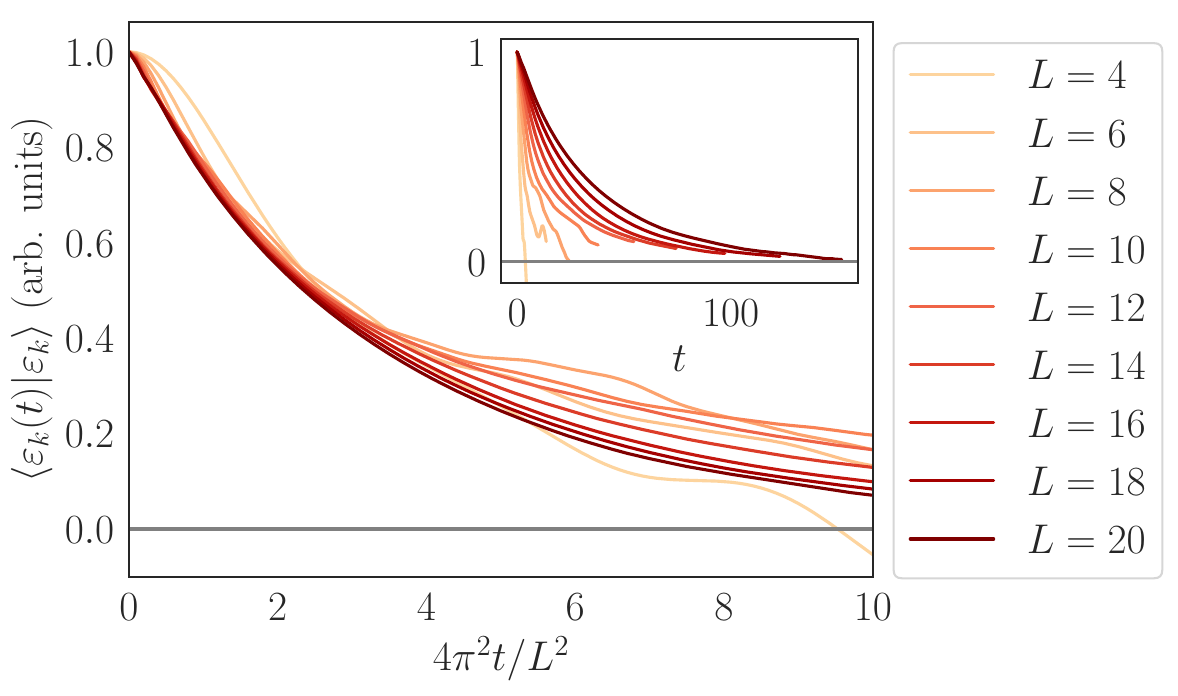}
  \end{minipage}

  \begin{minipage}{0.45\textwidth}
    \includegraphics[width=\textwidth]{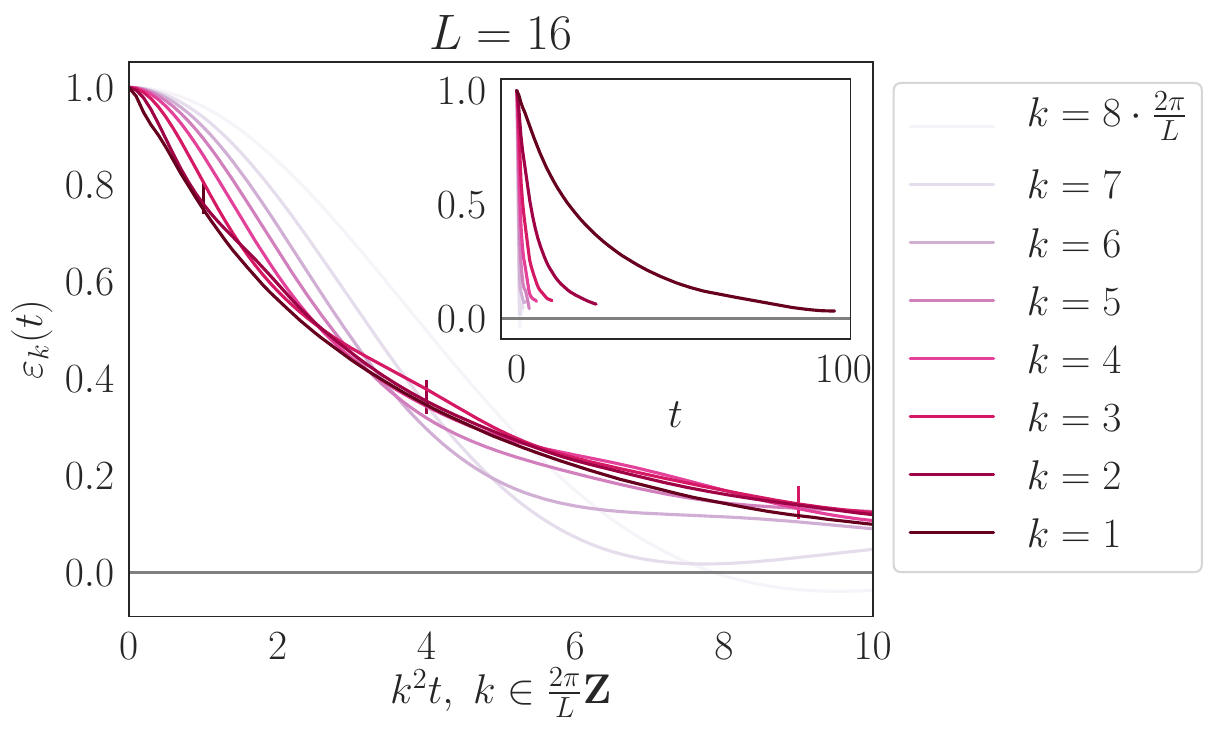}
  \end{minipage}

  \caption{
    \textbf{Fourier mode decay in full dynamics:} for the longest Fourier mode $k = \frac {2\pi}{L}$ as a function of system size $L$ \textbf{(top)} and as a function of Fourier mode $k$ for system size $L = 20$ \textbf{(bottom)}.
  }
  \label{fig:decay-exact-variables}
\end{figure}

\begin{figure}[t!]
  \begin{minipage}{0.45\textwidth}
    \includegraphics[width=\textwidth]{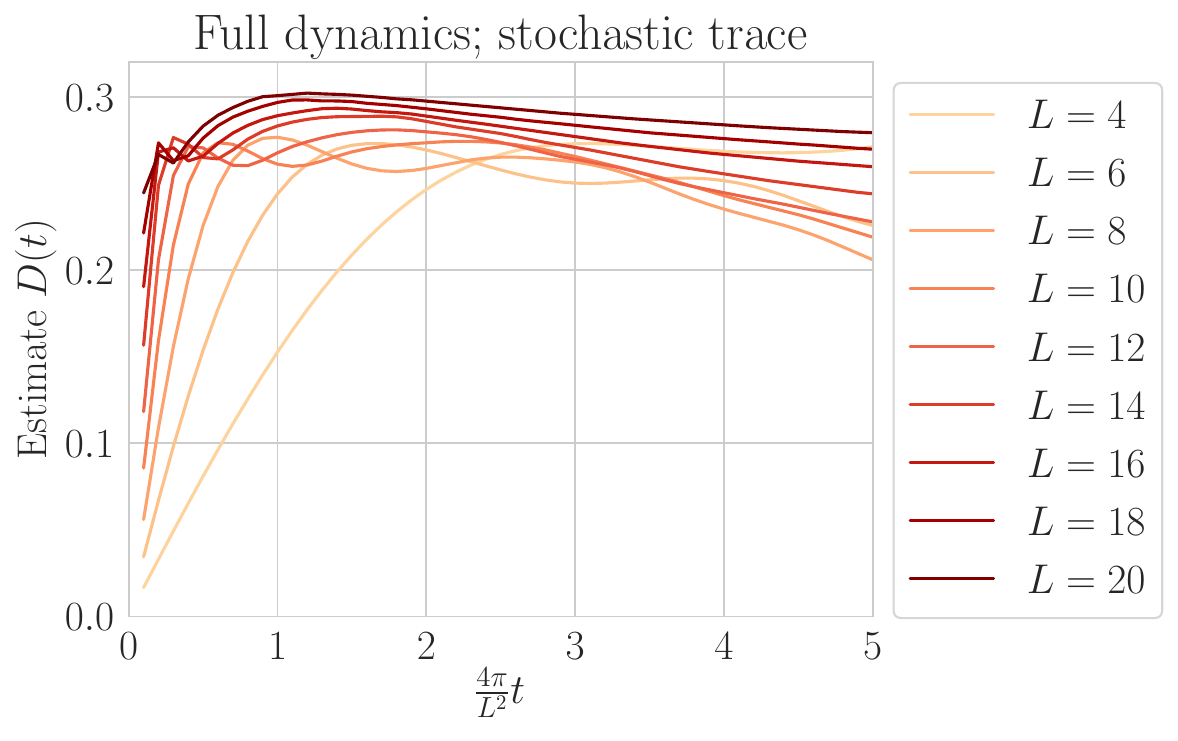}
  \end{minipage}

  \begin{minipage}{0.45\textwidth}
    \includegraphics[width=\textwidth]{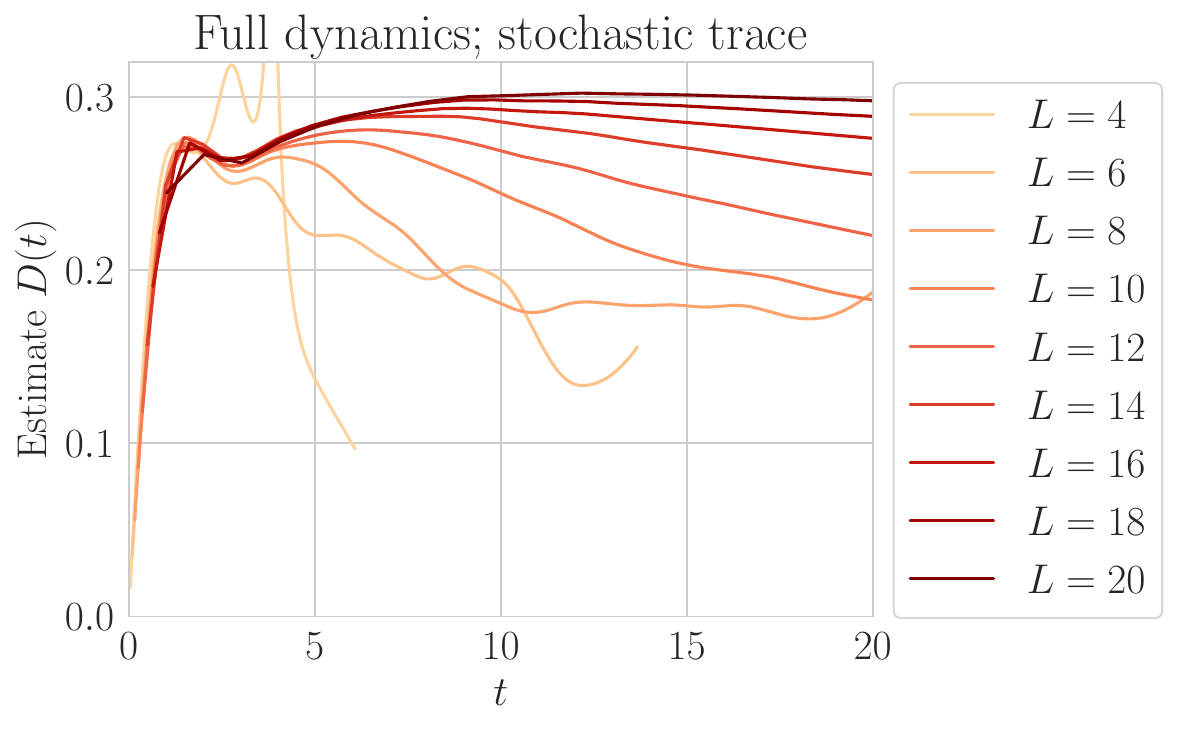}
  \end{minipage}

  \caption{
    \textbf{Time-varying estimate of diffusion coefficient}
    as a function of time and system size, with time rescaled by $\frac {4\pi}{L^2}$ (\textbf{top}) or unrescaled (\textbf{bottom}). 
  }
  \label{fig:diff-time}
\end{figure}

Diffusion of energy density is described by
\begin{align}
  \frac{\partial \varepsilon}{\partial t} = D \frac{\partial}{\partial x^2} \varepsilon\;;
\end{align}
for a Fourier mode
\begin{align}
  \label{eq:cosine-mode-app}
  \varepsilon_{k} = \sum_{j = 1}^L \cos(kj)
\end{align}
this gives
\begin{align}
  \label{eq:decay}
  \varepsilon_k(t) = e^{-Dk^2} \varepsilon_k\;.
\end{align}
I will ultimately show that the diffusion coefficient is $D \approx 0.3$.

Fig.~\ref{fig:decay-exact-variables} shows the decay
for the longest Fourier mode $k = \frac{2\pi}{L}$ as a function of length (top)
and as a function of wavenumber for $L = 20$ (bottom).

Consider first the behavior of the slowest Fourier mode
\begin{align}
  k = \frac {2\pi}{L}
\end{align}
as a function of system size (top plot).
One expects a characteristic timescale
\begin{align}
  t_{\mathrm{Th}} = \frac  {L^2}{4\pi^2D}\;,
\end{align}
so I rescale time by the diffusion coefficient agnostic factor $\frac{4\pi^2}{L^2}$.
For $L \ge 8$ and
\begin{align}
  t \lesssim 3 \frac{L^2t}{4\pi^2} \approx \frac{t}{L^2D4\pi^2 } = t_{\mathrm{Th}}\;
\end{align}
we see a good scaling collapse, improving as $L$ increases; this indicates that
the system is cleanly diffusive before its Thouless time---barring early-time transients,
which are obscured by the time rescaling.

Different Fourier modes for a fixed given system size also show a good diffusive scaling collapse, at least for $t \lesssim t_\Th$ (marked by vertical lines) and $k \lesssim 3\cdot\frac{2\pi}{L}$.
These $k$ correspond to length scales $\gtrsim 5$;
this is broadly consistent with the $L \gtrsim 8$ indicated by the scaling collapse in system size.

To clearly see the early-time non-diffusive behavior and estimate the diffusion coefficient,
write
\begin{align}
  D(t) = - \frac{t_{\mathrm{Th}}} {t}  \log\left(\frac{\tr[\varepsilon_k(t)\varepsilon_k]}{\tr[\varepsilon_k(0)\varepsilon_k(0)]}\right)\;.
\end{align}
(Note that here $k = {2\pi }/L$, hence the appearance of the Thouless time.)
Eq.~\eqref{eq:decay} predicts that this is a constant, always equal to the diffusion coefficient;
for systems that are not strictly diffusive we can treat this as a time-varying estimate of the diffusion coefficient.
Fig.~\ref{fig:diff-time} shows this $D(t)$.
In Fig.~\ref{fig:diff-time} bottom I do not rescale time.
There one can see that for
\begin{align}
  t \lesssim 5\;,
\end{align}
the system is non-diffusive.
For
\begin{align}
  1 \lesssim  \frac{4\pi}{L^2} t \lesssim 2
\end{align}
(and $L \gtrsim 8$) $D(t)$ is close to a constant,
indicating that the system is displaying diffusive decay of the Fourier mode in question.
I therefore estimate the diffusion coefficient by
\begin{align}
  D_{\mathrm{est}} = \mean_{ 1 \le \frac{4\pi}{L^2} t \le 2} D(t)
\end{align}
and the deviation from strictly diffusive behavior by
\begin{align}
  \Delta D_{\mathrm{est}} = \max_{ 1 \le \frac{4\pi}{L^2} t \le 2} D(t) - \min_{ 1 \le \frac{4\pi}{L^2} t \le 2}  D(t)\;.
\end{align}

\end{document}